\documentstyle[eqsecnum,epsf,preprint,tighten,aps]{revtex}
\newcommand{\vecz}[1]{\vec {#1} \,{}^2}
\begin{document}

\draft

\preprint{UGI-98-21}

\title{On QCD sum rules for vector mesons 
in nuclear medium\thanks{Work supported by GSI Darmstadt and BMBF.}}

\author{Stefan Leupold and Ulrich Mosel }
\address{Institut f\"ur Theoretische Physik, Universit\"at
Giessen,\\
D-35392 Giessen, Germany}

\date{May 13, 1998}

\maketitle

\begin{abstract}
Vector mesons show up in the electromagnetic current-current correlator. 
QCD sum rules provide a constraint on hadronic models for this correlator. 
This constraint is discussed for the case of finite nuclear density concerning
the longitudinal as well as the transverse part of the current-current correlator
at finite three-momentum. 
\end{abstract}

\section{Introduction}

In the last few years a lot of work has been investigated to study the behavior
of vector mesons in a medium with finite baryonic density. The basic motivation was
to find a sign of chiral symmetry restauration in heavy-ion collision experiments, 
when studying the dilepton spectra which correspond to the vector mesons. 
Indeed, the CERES experiments for S-Au and Pb-Au collisions show a novel feature
when compared to proton-nucleus collisions, namely an  
enhancement of the dilepton yield for invariant masses somewhat below the vacuum 
mass of the $\rho$ meson \cite{ceres1,ceres2,ceres3}. Some years ago it was argued 
by Brown and Rho that such an enhancement might be due to the restoration of chiral 
symmetry \cite{brownrho}. In their model they assumed that the masses of the 
vector mesons should scale with the quark
condensate, i.e.~drop with rising baryonic density. If this is true, the $\rho$
peak in the dilepton spectrum would be shifted to lower invariant mass. This might
be an explanation for the observed enhancement of
the dilepton yield in that region \cite{cassing,ko,bratkov}. However, the idea of
chiral symmetry restauration alone without additional model assumptions does not
provide a unique picture. Other scenarios predict a rising $\rho$ mass based on 
the effect that the $\rho$
becomes degenerate with its chiral partner, the $a1$ meson \cite{pisarski}.

Besides the problem what the consequences of chiral symmetry restauration might be
there is still the possibility that the experimental finding of the enhancement
might also be explained by conventional hadronic degrees of freedom. To clarify 
that issue hadronic models for the in-medium behavior of vector mesons were 
developed by various groups, see
e.g.~\cite{chanfray,herrmann,asakawa,rapp1,friman,rapp2,klingl97,eletsky,%
peters,kondrat98}. 
Some of them predict a large peak broadening of the $\rho$ meson or even distinct
new peak structures. It was found that the enhancement in the dilepton yield might 
also be explained
within a purely hadronic scenario, if a lot of strength is shifted to 
lower invariant mass \cite{rapp2,brat98}. So far one is not in a position to
confirm or rule out such hadronic models by experimental data. Therefore, it
is of interest to find additional model independent consistency checks which
should be obeyed by arbitrary hadronic models describing vector mesons and 
their in-medium
behavior. Such a consistency check is provided by the QCD sum rule approach. 

Originally, QCD sum rules were developed for vacuum processes not as a consistency 
check for hadronic models, but as an alternative
to them, i.e.~to deduce model independent information about hadrons from the
underlying quark and gluon degrees of freedom 
(see e.g.~\cite{shif79,reinders85,leinw}). 
An important ingredient for the description of vector mesons by QCD sum rules
is the assumption that the spectral functions of these resonances can be reliably 
approximated by $\delta$-functions (narrow width approximation). 
In this way, the experimentally found 
vector meson masses can be reproduced reasonably well. Of course, it is supported
by experiments that e.g.~the respective width of the spectral function of $\rho$ 
and $\omega$ meson is small as compared to the mass of the meson. However, it
is important to realize that the narrow width approximation is not a result, but
an ingredient of the traditional QCD sum rule approach. 

In the last few years QCD
sum rules were also developed for in-medium situations, i.e.~for hadronic matter
at finite temperature (see e.g.~\cite{hats93,adami}) or finite baryonic
density (e.g.~\cite{hats92,hats95,cohen95,jinlein,lee97}). For $\rho$ and
$\omega$ mesons it was found that their masses decrease with increasing temperature
and/or density, if the narrow width approximation is used for the parametrization
of the respective spectral function. In contrast to the vacuum case, there is,
however, no experimental support that the narrow width approximation is a reasonable
assumption for the in-medium case. Unfortunately, this approximation crucially 
influences the QCD sum rule prediction for a possible mass shift in a nuclear
surrounding. If a spectral function with an appropriately chosen large width 
is used in the QCD sum rule approach at finite density, one could get an unshifted 
meson mass, in
contrast to the finding utilizing the narrow width approximation. This was
first discussed in \cite{klingl97} using a specific hadronic model
and later systematically studied in \cite{leupold97}. 

This shows 
that QCD sum rules provide {\it no} 
model independent prediction about a possible mass shift of vector mesons in 
nuclear medium. 
Only together with some additional assumptions (e.g.~about the width of
the respective vector meson) a statement about the density dependence of the
masses of the vector mesons can be deduced from the sum rule analysis. 
Nevertheless, once a hadronic model for vector mesons has been chosen the sum rules
can be used as a consistency check for this model. We believe it to be important
to have such consistency checks, since it is not clear {\it a priori} if a 
hadronic model --- e.g.~with coupling constants derived from vacuum processes ---
yields a correct description of the in-medium behavior. 

In most of the studies on QCD sum rules only the vector mesons which are
at rest with respect to the medium were considered. On the other hand, since
Lorentz invariance is broken, the behavior of vector mesons clearly depends on
their velocity with respect to the surrounding. Indeed, some of the hadronic
models mentioned above yield very different spectral functions for different
three-momenta of the respective vector meson and for different polarizations
(e.g.~\cite{peters}). Only recently, the influence of finite three-momentum
on the QCD sum rule prediction for vector meson masses (within the classical 
narrow width approximation) was explored \cite{lee97}. Also at finite
three-momentum, of course, the narrow width approximation is an assumption which 
may be justified or not, but in any case it is not a model-independent statement. 

In this work we will derive QCD sum rules for $\rho$ and $\omega$ mesons for
arbitrary three-momentum of the respective vector meson with respect to the 
nuclear medium. The purpose is to provide a consistency check for hadronic models.
This is in contrast to the traditional QCD sum rule approach which aims at a
prediction for the vector meson mass assuming that the width of the vector meson 
is negligibly small (cf.~especially \cite{lee97} 
concerning the extension to non-vanishing three-momentum within the traditional 
QCD sum rule approach). In the application of the traditional QCD sum rule approach
to nuclear matter the attention was focused on the utilization of the sum rule
within the narrow width approximation \cite{hats92,hats95,jinlein,lee97}, 
rather than on a detailed discussion of the derivation of the sum rule and of
the calculation of the various condensates which contribute.
In the present article we try to bridge this gap. 

In the next section we introduce the basic quantity of interest, namely the
current-current correlator and present a dispersion relation which connects
the calculations for this correlator using hadronic degrees of freedom on the one 
hand side and quarks and gluons on the other. 
In section \ref{sec:ope} we sketch the method of operator product expansion
and calculate the  current-current correlator within that framework. 
In section \ref{sec:sumrule} we present a QCD sum rule derived from the dispersion
relation mentioned above. 
To get more insight into 
the various contributions calculated in section \ref{sec:ope}
we discuss in section \ref{sec:lindens} an approximation linear in the nuclear 
density. In section \ref{sec:crit} we discuss the various approximations which have
led to the results presented in the preceding sections. Finally we summarize our 
results in section \ref{sec:summary}.

\section{The current-current correlator}

The quantity we study in the following is the covariant 
time ordered current-current
correlator 
\begin{equation}
\Pi_{\mu\nu}(q) = i \int\!\! d^4\!x \, e^{iqx} \langle T j_\mu(x) j_\nu(0) 
\rangle 
\,.
  \label{eq:curcur}
\end{equation}
Here $j_\mu$ is an electromagnetic current with the isospin quantum number 
of the respective vector meson,
\begin{equation}
\label{eq:curud}
j_\mu = {1\over 2} \left( \bar u \gamma_\mu u \mp \bar d \gamma_\mu d \right)   
\end{equation}
where $-$ is for the $\rho$ meson and $+$ for the $\omega$. The current-current 
correlator enters e.g.~the cross section of $e^+ e^- \to $ hadrons 
(see e.g.~\cite{klingl97}). Within a simple vector meson dominance (VMD) picture 
the current (\ref{eq:curud})
can be identified with the vector meson which carries the respective isospin, 
e.g.~for the $\rho$ meson \cite{gale91}:
\begin{equation}
  \label{eq:vmd}
j_\mu \stackrel{\rm VMD}{=} {m_\rho^2 \over g_\rho } \rho_\mu
\end{equation}
where $\rho_\mu$ denotes the $\rho$ meson field amplitude
and $g_\rho$ the coupling of the $\rho$ meson to pions. 
Therefore, within simple VMD the current-current correlator is proportional to
the propagator of the respective vector meson. Speaking more generally, i.e.~without
referring to simple VMD, the vector meson propagator is closely related to the
current-current correlator. 

The expectation 
value in (\ref{eq:curcur}) is taken with respect to the surrounding medium. 
We study here a (isospin neutral) homogeneous equilibrated medium with finite 
nuclear density and vanishing temperature. 
In the medium Lorentz invariance is broken. All the formulae we
will present in the following refer to the Lorentz frame where the medium is at
rest, i.e.~where the spatial components of the baryonic current vanish. 

The current-current correlator can be decomposed in the following way \cite{gale91}:
\begin{equation}
  \Pi_{\mu\nu}(q) = \Pi_T(q) T_{\mu\nu}(q) + \Pi_L(q) L_{\mu\nu}(q) \,,
\end{equation}
where we have introduced two independent projectors $L_{\mu\nu}(q)$ and 
$T_{\mu\nu}(q)$ which both satisfy current conservation 
$q^\mu L_{\mu\nu}(q) = q^\mu T_{\mu\nu}(q) =0$ and add up to 
\begin{equation}
\label{eq:unitsub}
  T_{\mu\nu}(q) + L_{\mu\nu}(q) = g_{\mu\nu}-{q_\mu q_\nu \over q^2}  \,.
\end{equation}
The tensors $T$ and $L$ are transverse and longitudinal with respect to 
three-momentum $\vec q$, respectively. $T$ is given by 
\begin{equation}
\label{eq:deftrproj}
  T_{\mu\nu}(q) = \left\{ 
    \begin{array}{ccl}
      0 & , & \mu = 0 \; {\rm or} \; \nu =0 \,, \\
      -\delta^{ij} + {\displaystyle q^i q^j \over \displaystyle \vecz{q} }
& , & (\mu,\nu)=(i,j)  \,,
    \end{array}
\right.   
\end{equation}
while $L$ can be deduced from (\ref{eq:unitsub}). The scalar functions $\Pi_{T,L}$
can be obtained by 
\begin{equation}
  \label{eq:pit}
\Pi_T(q^2,\vecz{q}) = {1 \over 2} \Pi_{\mu\nu}(q) T^{\mu\nu}(q) 
= {1 \over 2} \left( 
\Pi^\mu_{\phantom{\mu}\mu} + {q^2 \over \vecz{q}} \Pi_{00}
\right)
\end{equation}
and
\begin{equation}
  \label{eq:pil}
\Pi_L(q^2,\vecz{q}) = \Pi_{\mu\nu}(q) L^{\mu\nu}(q) 
= -{q^2 \over \vecz{q}} \Pi_{00}     \,.
\end{equation}
To get the respective last equality in the last two equations, use is made of the
fact that $\Pi_{\mu\nu}(q)$ is a conserved quantity, i.e.~transverse with respect
to the {\it four}-momentum $q$. 
Note that $\Pi_T$ and $\Pi_L$ depend only on the invariant mass squared, $q^2$, and 
on the three-momentum squared, $\vecz{q}$. 
The latter property is due to the remaining $O(3)$
symmetry of the equilibrated system. 

At vanishing temperature the scalar functions $\Pi_T$, $\Pi_L$ deduced from the 
time ordered current-current correlator 
(\ref{eq:curcur}) can be related to the commutator (spectral function)
\begin{equation}
  \label{eq:defspecmunu}
{\cal A}_{\mu\nu}(q) := 
- {1\over 2} \int\!\! d^4\!x \, e^{iqx} \langle [j_\mu(x), j_\nu(0)] \rangle 
\end{equation}
in the following way (see e.g.~\cite{KB}):
\begin{equation}
  \label{eq:impia}
{\rm Im}\Pi_{T,L}(q^2,\vecz{q}) =  - {\rm sgn}(q_0) {\cal A}_{T,L}(q_0,\vecz{q})  
\end{equation}
where ${\cal A}_T$ and ${\cal A}_L$ are deduced from ${\cal A}_{\mu\nu}$ 
analogously to (\ref{eq:pit}) and (\ref{eq:pil}), respectively. At first sight,
it seems that ${\rm Im}\Pi_{T,L}$ does not only depend on $q^2$ and $\vecz{q}$ as
claimed above, but also on the sign of $q_0$. However, it is easy to 
check from the definition (\ref{eq:defspecmunu}) and the symmetry properties of the
system under consideration that ${\cal A}_T$ and ${\cal A}_L$ are antisymmetric 
with respect to the transformation $q_0 \to -q_0$. Therefore we define 
\begin{equation}
  \label{eq:specplus}
{\cal A}^+_{T,L}(q^2,\vecz{q}) := {\rm sgn}(q_0) {\cal A}_{T,L}(q_0,\vecz{q}) \,.
\end{equation}
Inserting this relation in (\ref{eq:impia}),
\begin{equation}
  \label{eq:impiapl}
{\rm Im}\Pi_{T,L}(q^2,\vecz{q}) =  - {\cal A}^+_{T,L}(q^2,\vecz{q})  \,,
\end{equation}
it becomes obvious that the dependence on the sign of $q_0$ is only apparent. 

For $q^2 \ll 0$ the current-current correlator (\ref{eq:curcur}) can be calculated
using Wilson's operator product expansion (OPE) \cite{wilson69} 
for quark and gluonic degrees of freedom \cite{shif79} (for in-medium calculations
see e.g.~\cite{hats93,hats92,hats95}). In the following we shall call the result 
of that calculation $\Pi^{{\rm OPE}}_{T,L}$. 
On the other hand, a hadronic model (e.g.~for vector mesons 
\cite{chanfray,herrmann,asakawa,rapp1,friman,rapp2,klingl97,peters}
using one or the other form of VMD) can give an expression 
for the current-current correlator valid in the time like region $q^2 > 0$. We 
denote the result of the hadronic model by $\Pi^{{\rm HAD}}_{T,L}$. 
A second representation in the space like region which has to match 
$\Pi^{{\rm OPE}}_{T,L}$ can be obtained from $\Pi^{{\rm HAD}}_{T,L}$ by
utilizing a twice subtracted dispersion relation. We find 
\begin{equation}
  \label{eq:disp1}
\Pi_{T,L}(q^2,\vec q\,^2) = \Pi_{T,L}(0,\vec q\,^2) + c_{T,L}(\vec q\,^2) q^2 
- {q^4 \over \pi} \int\limits_{-\vecz{q}}^\infty \!\! ds \,
{{\cal A}^+_{T,L}(s,\vecz{q}) \over (s-q^2-i\epsilon)(s+i\epsilon')^2} 
\end{equation}
with the subtraction constant
\begin{equation}
  \label{eq:unimp1}
c_{T,L}(\vec q\,^2) = 
\left. 
{\partial \Pi_{T,L}(k^2,\vecz{q}) \over \partial (k^2) } 
\right\vert_{k^2 = 0}    \,.
\end{equation}
As we shall see below ${\cal A}^{{\rm HAD}}_{T,L}(q)$ diverges linearly with 
$q^2$. Therefore we have used above a twice subtracted dispersion relation.
In the space like region for $Q^2:= -q^2 \gg 0$ we get the following connection
between the current-current correlator calculated from OPE on the one hand side
and from a hadronic model on the other:
\begin{equation}
  \label{eq:opehad}
\Pi^{{\rm OPE}}_{T,L}(Q^2,\vec q\,^2) = 
\Pi^{{\rm HAD}}_{T,L}(0,\vec q\,^2) - c_{T,L}(\vec q\,^2) Q^2 
+ {Q^4 \over \pi} \int\limits_{-\vecz{q}}^\infty \!\! ds \,
{{\rm Im}\Pi^{{\rm HAD}}_{T,L}(s,\vecz{q}) \over (s+Q^2-i\epsilon)(s+i\epsilon')^2} 
\end{equation}
where we have used (\ref{eq:impiapl}) to express the spectral function in terms of 
the imaginary part of $\Pi$. 

In the next section we shall elaborate on the calculation of the l.h.s.~of 
(\ref{eq:opehad}).

\section{Operator product expansion} \label{sec:ope}

Within the method of operator product expansion
we have to calculate the current-current correlator 
(\ref{eq:curcur}) for large space like momenta $Q^2=-q^2$. Here the relevant length
scale for the $x$-integration given by the inverse of $1/\sqrt{Q^2}$ is small. 
This defines the hard scale in our problem. 
Suppose that the distance $x$ is much smaller than the typical length
of the system (soft scale). 
The latter might be characterized e.g.~by the average particle distance in the 
medium or $1/\Lambda_{\rm QCD}$ as the scale
where non perturbative effects appear. If $x$ is that small 
it is reasonable to assume that a product of local operators $A$ and $B$, i.e.
\begin{equation}
  \label{eq:nolo}
A(x) B(0)  \,,
\end{equation}
should look like a local operator,
since the system cannot resolve such small distances $x$. Thus we find
\begin{equation}
  \label{eq:wilsonexp}
A(x) B(0)  \approx  \sum\limits_n C_n(x) \, {\cal O}_n   \,,
\end{equation}
where $C_n$ denotes c-number functions (Wilson coefficients) 
and ${\cal O}_n$ {\it local} operators. 
The only dependence on the system under consideration enters via the respective
matrix elements of the local operators ${\cal O}_n$. Thus, the dependence on the
soft scale is entirely given by the local operators. On the other hand, the Wilson 
coefficients $C_n$ can be calculated independently from the system under 
consideration. Since the operators ${\cal O}_n$ are local, the dependence on the
hard scale (here given by the $x$-dependence) enters only the Wilson coefficients. 
Thus we have achieved a separation of the hard from the soft scale. 

In our case, the operators $A$ and $B$ are the currents $j_\mu$ and $j_\nu$, 
respectively. 
The Fourier transformation which appears in (\ref{eq:curcur}) does not touch
the local operators ${\cal O}_n$, but only changes the $x$-dependence of $C_n$
into a $q$-dependence. From the line of arguments one can already guess that 
the Wilson coefficients finally yield a power series in $1/Q^2$ (corrected by
logarithms from renormalization). The expectation values of the local operators
${\cal O}_n$ (condensates) show up as coefficients of that series. 
On dimensional grounds it is obvious that the higher the dimension 
(in terms of masses) of a condensate is, the more
it is suppressed by powers of $1/Q^2$. 

In the following we will consider condensates up to dimension 6. In vacuum only
scalar condensates contribute. For the case at hand, however, 
the condensates might also carry spin, since Lorentz invariance
is broken. It is common practice to classify the condensates by their dimensionality
$d$ and their twist $\tau$. The latter is defined as the difference of dimension
$d$ and spin $s$, i.e.~$\tau = d - s$. We decompose the current-current correlator
(\ref{eq:curcur}) in the following way:
\begin{equation}
  \label{eq:curcurdtau}
\Pi^{{\rm OPE}}_{\mu\nu} \approx  \Pi^{{\rm scalar}}_{\mu\nu} + 
\Pi^{d=4,\tau=2}_{\mu\nu} + \Pi^{d=6,\tau=2}_{\mu\nu} + 
\Pi^{d=6,\tau=4}_{\mu\nu} 
\end{equation}
where we have neglected contributions from higher dimensional condensates. 
We will discuss the various contributions separately in the following subsections.

\subsection{Scalar condensates}  \label{sec:scal}

The contribution of the scalar condensates to the current-current correlator
for the system with finite nuclear density is formally identical to the vacuum case.
The only difference is that in the former case the expectation value is taken 
with respect to the medium. The latter case was already discussed in the original
paper by Shifman et al.~\cite{shif79}. The result is (cf.~e.g.~\cite{hats93} for 
details)
\begin{equation}
  \label{eq:scalope1}
\Pi^{{\rm scalar}}_{\mu\nu}(q) = 
\left( g_{\mu\nu}-{q_\mu q_\nu \over q^2} \right) 
Q^2 R^{{\rm scalar}}(Q^2) 
\end{equation}
with
\begin{eqnarray}
R^{{\rm scalar}}(Q^2) & \approx &
-{1 \over 8 \pi^2} \left( 1+ {\alpha_s \over \pi} \right) 
  {\rm ln}\left(Q^2\over \mu^2 \right) 
+ {m_q \over 2 Q^4} \left\langle \bar u u + \bar d d \right\rangle 
+ {1 \over 24 Q^4} \left\langle {\alpha_s \over \pi} G^2 \right\rangle 
\nonumber \\
&& {} - {\pi \alpha_s \over 2 Q^6} \left\langle 
(\bar u \gamma_\mu \gamma_5 \lambda^a u \mp \bar d \gamma_\mu \gamma_5 \lambda^a d)
(\bar u \gamma^\mu \gamma_5 \lambda^a u \mp \bar d \gamma^\mu \gamma_5 \lambda^a d)
\right\rangle
\nonumber \\
&& {} - {\pi \alpha_s \over 9 Q^6} \left\langle
(\bar u \gamma_\mu \lambda^a u + \bar d \gamma_\mu \lambda^a d)
\sum\limits_{q = u,d,s} \bar q \gamma^\mu \lambda^a q 
\right\rangle   \,.
  \label{eq:scal2}
\end{eqnarray}
where we have neglected contributions quadratic in the light current quark mass
$m_q$ as well as differences in the masses of up and down quark. This is reasonable,
since the hard scale $\sqrt{Q^2}$ is typically of the order of $1\,$GeV 
(the order of magnitude of the considered vector meson),
while the masses of up and down quarks are of the order of a few MeV. Again the 
$-$ sign refers to the $\rho$ meson and the $+$ to the $\omega$. To
simplify (\ref{eq:scal2}) we assume that the quark condensates of up and down
quarks are approximately the same. Furthermore, we replace the four-quark 
condensates by products of two-quark condensates. Since it is not clear how 
accurate the assumption of ground state saturation (Hartree approximation) is, 
we multiply the result with a (still to be determined) factor $\kappa$. 
We end up with 
\begin{eqnarray}
R^{{\rm scalar}}(Q^2) & \approx &
-{1 \over 8 \pi^2} \left( 1+ {\alpha_s \over \pi} \right) 
  {\rm ln}\left(Q^2\over \mu^2 \right) 
+ {m_q \over Q^4} \left\langle \bar q q \right\rangle 
+ {1 \over 24 Q^4} \left\langle {\alpha_s \over \pi} G^2 \right\rangle 
- {112 \pi \alpha_s \over 81 Q^6} \kappa 
\left\langle \bar q q \right\rangle^2   \,.
  \label{eq:scal3}
\end{eqnarray}
Note that there is no difference between $\rho$ and $\omega$ any more, since
there are no terms like $\langle \bar u d \rangle$ in an isospin neutral medium. 

Of course, the crucial question is, how to evaluate the expectation values
with respect to the nuclear medium. If the density is small, it is reasonable
to approximate the medium by a Fermi gas of free nucleons, i.e.,
\begin{equation}
  \label{eq:fermi}
\langle {\cal O} \rangle \approx \langle {\cal O} \rangle_0 \,\, + \,\, 
4 \int\limits_{\vert \vec k \vert \le k_F } {d^3 k \over (2\pi)^3 \,2 E_k} 
\langle N(\vec k) \vert {\cal O} \vert N(\vec k) \rangle  \,,
\end{equation}
where $\langle {\cal O} \rangle_0 $ denotes the vacuum expectation value of an
arbitrary operator ${\cal O}$, $k_F$ the Fermi momentum, 
$E_k=\sqrt{m_N^2+\vec k\,^2}$ the energy of a nucleon, and
$\vert N(\vec k) \rangle$ a single (isospin averaged) nucleon state with momentum
$\vec k$ normalized according to
\begin{equation}
  \label{eq:normnuc}
\langle N(\vec k) \vert N(\vec k') \rangle 
= (2\pi)^3 \, 2 E_k \, \delta(\vec k - \vec k')   \,.
\end{equation}
We will use the approximation (\ref{eq:fermi}) throughout this work and comment 
on it in section \ref{sec:crit}. 

If ${\cal O}$ is a scalar operator, the expectation
value $\langle N(\vec k) \vert {\cal O} \vert N(\vec k) \rangle $ is a scalar as
well and therefore independent of the momentum of the nucleon. Thus we get
\begin{equation}
  \label{eq:condscal}
\langle {\cal O}_{\rm scalar} \rangle \approx 
\langle {\cal O}_{\rm scalar} \rangle_0 + 
4 \,\langle N(0) \vert {\cal O}_{\rm scalar} \vert N(0) \rangle 
\int\limits_{\vert \vec k \vert \le k_F } {d^3 k \over (2\pi)^3 \, 2 E_k}   \,.
\end{equation}
For the evaluation of the condensates in (\ref{eq:scal3}) we need to know the 
expectation values of the quark and gluon condensate with respect to single
nucleon states. The former can be related to the nucleon sigma term 
\cite{hats92},\footnote{Note that the normalization of the nucleon state in 
\cite{hats92} is different from ours.}
\begin{equation}
  \label{eq:nucsig}
\langle N(0) \vert \bar q q \vert N(0) \rangle = {m_N \sigma_N \over m_q}  \,,
\end{equation}
while the latter can be calculated from the trace anomaly of QCD,
\begin{equation}
  \label{eq:traceano}
\left\langle N(0) \left\vert 
{\alpha_s \over \pi} G^2 
\right\vert N(0) \right\rangle =
- {16 \over 9} m_N m_N^{(0)}  \,.
\end{equation}
Here $\sigma_N$ denotes the nucleon sigma term and $m_N^{(0)}$ the nucleon mass in 
the chiral limit. 

Finally, we have to make a choice for the parameter $\kappa$ which parametrized
the deviation of the four-quark condensate from the product of two-quark 
condensates. Even for the vacuum case, the question about the value for $\kappa$ is
not settled yet (cf.~e.g.~\cite{hats92,hats93,hats95,klingl97,leinw}).
In addition, $\kappa$ might depend on the nuclear density. For simplicity, we
take in the following the vacuum value for $\kappa$ also for arbitrary finite 
densities, but note that this introduces an uncertainty into the evaluation
of the OPE.
All the parameters not specified so far are taken from \cite{klingl97}
and listed in table \ref{tab:para}. We discuss this choice in section 
\ref{sec:crit}. 

Using (\ref{eq:pit},\ref{eq:pil},\ref{eq:scalope1},\ref{eq:scal3},%
\ref{eq:condscal}-\ref{eq:traceano}) we are able to calculate the contribution
of the scalar condensates to the l.h.s.~of \ref{eq:opehad}). In the next subsection
we discuss the contribution of the twist-2 condensates with $d$$=$4.

\subsection{Twist-2 spin-2 condensates} \label{sec:t2s2}

In vacuum only scalar condensates contribute to the current-current correlator
since there is no Lorentz vector which can account for the spin of a
non-scalar condensate. Contrary to the vacuum case, in a nuclear medium the 
baryonic current can
yield the spin. Using the approximation (\ref{eq:fermi}) we find for spin-2
condensates
\begin{equation}
  \label{eq:fermispin2}
\langle {\cal O}_{\mu\nu} \rangle \approx 
4 \int\limits_{\vert \vec k \vert \le k_F } {d^3 k \over (2\pi)^3 \,2 E_k} 
\langle N(\vec k) \vert {\cal O}_{\mu\nu} \vert N(\vec k) \rangle  \,.
\end{equation}
In this approximation the four-momentum of the nucleon accounts for the spin
of the condensate. Thus we get
\begin{equation}
  \label{eq:trls}
\langle N(\vec k) \vert {\cal O}_{\mu\nu} \vert N(\vec k) \rangle \sim 
k_\mu k_\nu - {g_{\mu\nu} \over 4} m_N^2 =: S_ {\mu\nu}(k)  \,.
\end{equation}
Note that the non-scalar operators are traceless with respect to the Lorentz
indices. 

Expectation values of twist-2 condensates with respect to single nucleon states
as they appear in (\ref{eq:fermispin2}) are thoroughly studied
in deep inelastic scattering (DIS), 
albeit for a somewhat different kinematical situation. 
We can utilize the results obtained there for our case at hand --- as already
pointed out in \cite{hats92,hats95,lee97}. Therefore, we will not give a detailed
calculation for the contributions of these condensates to the current-current
correlator, but instead present a recipe how to deduce the necessary information
from the DIS calculations of \cite{bard78,flor81,glu92}. 

The twist-2 operators of dimensionality 4 which contribute are given by 
\begin{equation}
  \label{eq:op22q}
{\cal ST} \,i( \bar u \gamma_\mu D_\nu u + \bar d \gamma_\mu D_\nu d )
\end{equation}
and
\begin{equation}
  \label{eq:op22g}
{\cal ST} \, G^\kappa_{\phantom{\kappa}\mu} G_{\kappa\nu}   \,.
\end{equation}
Here ${\cal ST}$ denotes an operator producing an expression which is
symmetric and traceless with respect to the Lorentz indices $\mu$ and $\nu$. 
$D_\nu$ is the covariant
derivative and $G_{\kappa\nu}$ the 
gluonic field strength tensor. In principle, composite operators mix under the
renormalization group, if they have the same quantum numbers \cite{pastar}. 
To study that mixing we have to decompose (\ref{eq:op22q}) in a flavor singlet
part which mixes with the gluonic operator (\ref{eq:op22g}) and a flavor
non-singlet part which does not mix. For the energy region of interest, 
i.e.~roughly about the masses of $\rho$ and $\omega$ meson, we have to deal with
three active flavors. Therefore, we decompose (\ref{eq:op22q}) according to
\begin{equation}
  \label{eq:decflav}
u + d = {1\over 3} \left[ 2 \, (u+d+s) + (u+d-2s) \right]  \,,
\end{equation}
where $u$ is an abbreviation for $\bar u \gamma_\mu D_\nu u$, etc. Renormalization
group mixing applies to $(u+d+s)$ and $G$, i.e.~schematically
\begin{equation}
  \label{eq:rgm}
u+d \to {1\over 3} \left[ 2 \, (u+d+s+G) + (u+d-2s) \right] 
= u + d + {2\over 3} G  \,.
\end{equation}

The contribution of the twist-2 spin-2 condensates to the current-current correlator
(\ref{eq:curcur}) can be written as 
\begin{equation}
  \label{eq:scattintro}
\Pi^{d=4,\tau=2}_{\mu\nu}(q) 
= 4 \int\limits_{\vert \vec k \vert \le k_F } {d^3 k \over (2\pi)^3 \,2 E_k} 
\left({1\over 2} \right)^2
\sum\limits_{\psi = u,d} T_{\mu\nu}^{(\psi) \, s=2,\tau=2}(q,k)  
\end{equation}
with the twist-2 spin-2 contribution to the forward scattering amplitude between
a nucleon and a quark current,
\begin{equation}
  \label{eq:fsa22}
T_{\mu\nu}^{(\psi) \, s=2,\tau=2}(q,k) := 
i \int\!\! d^4\!x \, e^{iqx} 
\langle N(\vec k) \vert 
T \bar \psi(x) \gamma_\mu \psi(x) \, \bar \psi(0) \gamma_\nu \psi(0) 
\vert N(\vec k) \rangle_{s=2,\tau=2}   \,.
\end{equation}
The latter quantity is calculated in \cite{bard78} for the DIS case. 
Note that the factor $1/2$ in the definition of the current (\ref{eq:curud}) 
--- which enters (\ref{eq:curcur}) quadratically --- is not contained in the
definition of the forward scattering amplitude, but is given in 
(\ref{eq:scattintro}) explicitly. 

The strategy to utilize the DIS results for the forward scattering amplitude 
(\ref{eq:fsa22}) is the following: We make a general ansatz for this amplitude,
specify it to the DIS case, match it with the calculations
of \cite{bard78}, and determine in this way the unknown quantities of the general 
ansatz. 

The Lorentz structure of the forward scattering amplitude 
(\ref{eq:fsa22}) must be built up from the tensor $g_{\mu\nu}$, the four-momentum
$q$ of the quark current, and the tensor $S_{\mu\nu}(k)$, 
defined in (\ref{eq:trls}).
In addition, the amplitude has to obey current conservation. Finally, it must be
symmetric with respect to an exchange of the Lorentz indices. A general ansatz
which fulfills all requirements is
\begin{eqnarray}
T_{\mu\nu}^{(\psi) \, s=2,\tau=2}(q,k) & = & 
B_1 \, [q^4 S_{\mu\nu}(k) - q^2 q_\mu q^\alpha S_{\nu\alpha}(k) 
- q^2 q_\nu q^\alpha S_{\mu\alpha}(k) 
+ g_{\mu\nu} q^2 q^\alpha q^\beta S_{\alpha\beta}(k)  ]
\nonumber \\
&& {} + B_2 \, (q_\mu q_\nu - g_{\mu\nu} q^2) \, 
q^\alpha q^\beta S_{\alpha\beta}(k) 
  \label{eq:ans22}
\end{eqnarray}
with so far arbitrary coefficients $B_1$ and $B_2$ which might depend on $q^2$
and $q\cdot k$. 

The kinematical situation of DIS is such that both $-q^2$ and $q\cdot k$ are
large with the Bjorken variable $x=-q^2/( \,2q\cdot k)$ fixed. In this limit
only the $k_\mu k_\nu$ term of $S_{\mu\nu}(k)$ has to be taken into 
account.\footnote{Note that this is not true for the case we are actually 
interested in. Because the three-momentum $\vec q$ might be small,
we also have to take into account the $g_{\mu\nu}$ term. 
However, for the determination of the coefficients $B_1$ and $B_2$ this does not 
matter.}
We get
\begin{equation}
  \label{eq:scattampdis}
T_{\mu\nu}^{(\psi) \, s=2,\tau=2\, {\rm DIS}}(q,k) 
= - q^2 (q\cdot k)^2 (B_1 \, d_{\mu\nu} + B_2 \, e_{\mu\nu} )
\end{equation}
with the tensors \cite{bard78}
\begin{equation}
  \label{eq:bard1}
e_{\mu\nu} = g_{\mu\nu} - {q_\mu q_\nu \over q^2}
\end{equation}
and
\begin{equation}
  \label{eq:bard2}
d_{\mu\nu} = - {k_\mu k_\nu \over (q\cdot k)^2} 
+ {k_\mu q_\nu + k_\nu q_\mu \over q\cdot k } - g_{\mu\nu} \,.
\end{equation}
By comparison of (\ref{eq:scattampdis}) with equation (2.4) of \cite{bard78}
we find 
\begin{equation}
  \label{eq:detb1}
B_1 = -{4 \over q^6} \sum\limits_i C^i_{2,2} A_2^i 
\end{equation}
and
\begin{equation}
  \label{eq:detb2}
B_2 = -{4 \over q^6} \sum\limits_i C^i_{L,2} A_2^i   \,,
\end{equation}
with the process independent coefficient function $C^i_{r,n}$ and the $(n-1)$th 
moment $A_n^i$ of the distribution of the parton $i$ in the nucleon.
Expressions for the former can be found e.g.~in \cite{flor81} including 
$\alpha_s$ corrections.\footnote{Note that our notation (basically adopted from
\cite{bard78}) is somewhat different from the one of \cite{flor81}: Our coefficient
function $C^j_{2,n}$ is identical to $C^N_{2,j}$ of \cite{flor81} with $n=N$. The
longitudinal coefficient functions differ by a factor 2: $C^j_{L,n} = 2 C^N_{L,j}$,
where again the former is our coefficient function and the latter the one of 
\cite{flor81}.}
The latter is given by \cite{lee97}
\begin{equation}
  \label{eq:defmomq}
A^\psi_n = 2 \int\limits_0^1 \!\! dx \, x^{n-1} 
\left[ \psi(x,\mu^2) + \bar \psi(x,\mu^2) \right]  
\end{equation}
for quarks and 
\begin{equation}
  \label{eq:defmomg}
A^G_n = 2 \int\limits_0^1 \!\! dx \,x^{n-1} \, G(x,\mu^2) 
\end{equation}
for gluons. The parton distributions $\psi$, $\bar \psi$, and $G$ at the 
renormalization scale $\mu^2$ are parametrized in \cite{glu92}. 

Now we collect all the obtained information to get the 
contribution of the twist-2 spin-2 condensates to the current-current correlator:
\begin{equation}
  \label{eq:piscam22fin}
\Pi^{d=4,\tau=2}_{\mu\nu}(q) 
= \int\limits_{\vert \vec k \vert \le k_F } {d^3 k \over (2\pi)^3 \,2 E_k} 
T_{\mu\nu}^{(u+d) \, s=2,\tau=2}(q,k)  
\end{equation}
with
\begin{eqnarray}
T_{\mu\nu}^{(u+d) \, s=2,\tau=2}(q,k) & = & -{4 \over q^4} 
[q^2 S_{\mu\nu}(k) - q_\mu q^\alpha S_{\nu\alpha}(k) 
- q_\nu q^\alpha S_{\mu\alpha}(k) 
+ g_{\mu\nu} q^\alpha q^\beta S_{\alpha\beta}(k)  ]
\nonumber \\
&& \phantom{-} \times 
\left( C_{2,2}^q A_2^{u+d} + {2\over 3} C_{2,2}^G A_2^G \right)
\nonumber \\
&&
- {4\over q^6} (q_\mu q_\nu - g_{\mu\nu} q^2) \, 
q^\alpha q^\beta S_{\alpha\beta}(k) 
\left( C_{L,2}^q A_2^{u+d} + {2\over 3} C_{L,2}^G A_2^G \right)
  \label{eq:scam22fin}
\end{eqnarray}
where we have taken into account that the gluonic contribution enters with 
a factor $2/3$ according to (\ref{eq:rgm}). The coefficient functions and the
moments of the parton distributions are listed in table \ref{tab:coeff}. 
We note in passing that the momentum integrations in (\ref{eq:piscam22fin}) can be
performed analytically. Since it is not illuminating to present the lengthy result
of these integrations we stick to the compact form given by (\ref{eq:piscam22fin}) 
and (\ref{eq:scam22fin}).

\subsection{Twist-2 spin-4 condensates} \label{sec:t2s4}

The twist-2 spin-4 condensates can be treated in the same way as the twist-2 spin-2
condensates. We use the approximation (\ref{eq:fermi}) to find
\begin{equation}
  \label{eq:fermispin4}
\langle {\cal O}_{\mu\nu\kappa\lambda} \rangle \approx 
4 \int\limits_{\vert \vec k \vert \le k_F } {d^3 k \over (2\pi)^3 \,2 E_k} 
\langle N(\vec k) \vert {\cal O}_{\mu\nu\kappa\lambda} \vert N(\vec k) \rangle  \,.
\end{equation}
For the expectation value with respect to a single nucleon state we get the
decomposition
\begin{eqnarray}
\lefteqn{\langle N(\vec k) \vert {\cal O}_{\mu\nu\kappa\lambda} 
\vert N(\vec k) \rangle} \nonumber \\  
& \sim & 
k_\mu k_\nu k_\kappa k_\lambda 
- {1\over 8}( k_\mu k_\nu g_{\kappa\lambda} m_N^2 + \mbox{5 permutations})
+ {1\over 48} (g_{\mu\nu} g_{\kappa\lambda} m_N^4 + \mbox{2 permutations}) 
\nonumber \\ 
&=:&  S_{\mu\nu\kappa\lambda}(k)   \,.
  \label{eq:trlspin4}
\end{eqnarray}
The relevant operators are 
\begin{equation}
  \label{eq:op24q}
{\cal ST} \,i( \bar u \gamma_\mu D_\nu D_\kappa D_\lambda u 
+ \bar d \gamma_\mu D_\nu D_\kappa D_\lambda d )
\end{equation}
and
\begin{equation}
  \label{eq:op24g}
{\cal ST} \, G^\alpha_{\phantom{\alpha}\mu} D_\nu D_\kappa G_{\alpha\lambda}   \,.
\end{equation}
By comparison with the DIS calculations we find 
\begin{equation}
  \label{eq:piscam24fin}
\Pi^{d=6,\tau=2}_{\mu\nu}(q) 
= \int\limits_{\vert \vec k \vert \le k_F } {d^3 k \over (2\pi)^3 \,2 E_k} 
T_{\mu\nu}^{(u+d) \, s=4,\tau=2}(q,k)  
\end{equation}
with
\begin{eqnarray}
T_{\mu\nu}^{(u+d) \, s=4,\tau=2}(q,k) & = & -{16 \over q^{10}} 
(q_\mu q_\nu q^\alpha q^\beta - g_\mu^{\phantom{\mu} \alpha} q^2 q_\nu q^\beta
- g_\nu^{\phantom{\nu} \alpha} q^2 q_\mu q^\beta 
+ g_\mu^{\phantom{\mu} \alpha} g_\nu^{\phantom{\nu} \beta} q^4 ) \, 
q^\gamma q^\delta S_{\alpha \beta \gamma \delta}(k) 
\nonumber \\
&& \phantom{-} \times 
\left( C_{2,4}^q A_4^{u+d} + {2\over 3} C_{2,4}^G A_4^G \right)
\nonumber \\
&&
-{16 \over q^{10}} (q_\mu q_\nu - g_{\mu\nu} q^2) \, 
q^\alpha q^\beta q^\gamma q^\delta S_{\alpha \beta \gamma \delta}(k) 
\nonumber \\
&& \phantom{-} \times 
\left( (C_{L,4}^q - C_{2,4}^q) A_4^{u+d} 
+ {2\over 3} (C_{L,4}^G - C_{2,4}^G) A_4^G \right)  \,,
  \label{eq:scam24fin}
\end{eqnarray}
where the tensor $S_{\alpha \beta \gamma \delta}(k)$ is defined in 
(\ref{eq:trlspin4}) and the coefficient functions and parton distribution moments 
are listed in table \ref{tab:coeff}.

\subsection{Twist-4 spin-2 condensates}

Finally we turn to the higher twist condensates. For our calculation up to
dimensionality $d$$=$$6$ we need condensates with twist $\tau$$=$$4$ and 
spin $s$$=$$2$ (besides the higher twist condensates which are scalar and
already discussed in subsection \ref{sec:scal}). These condensates 
are \cite{shur81,jaffe81,choi93,lee94}
\begin{equation}
  \label{eq:condt41}
{\cal O}^1_{\mu\nu} := {\cal ST} \, g^2 \, {1\over 4}
(\bar u \gamma_\mu \gamma_5 \lambda^a u \mp \bar d \gamma_\mu \gamma_5 \lambda^a d)
(\bar u \gamma_\nu \gamma_5 \lambda^a u \mp \bar d \gamma_\nu \gamma_5 \lambda^a d)
\, ,
\end{equation}
\begin{equation}
  \label{eq:condt42}
{\cal O}^2_{\mu\nu} := {\cal ST} \, g^2 \, {1\over 4}
(\bar u \gamma_\mu \lambda^a u + \bar d \gamma_\mu \lambda^a d)
\sum\limits_{q = u,d,s} \bar q \gamma_\nu \lambda^a q   \,,
\end{equation}
\begin{equation}
  \label{eq:condt4g}
{\cal O}^g_{\mu\nu} := {\cal ST} \, ig \, {1\over 4}
( \bar u \, \{ D_\mu, \tilde G_{\nu\alpha} \} \gamma^\alpha \gamma_5 u
+ \bar d \, \{ D_\mu, \tilde G_{\nu\alpha} \} \gamma^\alpha \gamma_5 d )   \,,
\end{equation}
\begin{equation}
  \label{eq:condt4no}
{\cal ST} \, g \, {1\over 4}
( \bar u \, [D_\mu , G_{\nu\alpha}] \, \gamma^\alpha u 
+ \bar d \, [D_\mu , G_{\nu\alpha}] \, \gamma^\alpha d )   \,,
\end{equation}
and
\begin{equation}
  \label{eq:condt4m}
{\cal ST} {1\over 4}
(m_u \, \bar u \, D_\mu D_\nu u + m_d \, \bar d \, D_\mu D_\nu d )  \,,
\end{equation}
where the $-$ sign in (\ref{eq:condt41}) corresponds to the $\rho$ and the $+$ sign
to the $\omega$ meson. 
For an unknown reason the condensate given in (\ref{eq:condt4no}) actually 
does not contribute to the current-current correlator \cite{jaffe81}; 
we have listed it here for the sake of completeness, only. In the following
we will neglect the condensate (\ref{eq:condt4m}), since it is proportional to
the very small light quark masses and therefore suppressed. Additionally, we will
neglect the contribution of the strange quarks to the nucleon expectation values
of the operators given above \cite{choi93}. 

As in the previous subsections we use the Fermi gas approximation which
for the case of spin-2 condensates we have already given in (\ref{eq:fermispin2}).
We can also use the Lorentz decomposition from (\ref{eq:trls}): 
\begin{equation}
  \label{eq:trltwist4}
\langle N(\vec k) \vert {\cal O}^j_{\mu\nu} \vert N(\vec k) \rangle 
=: A^j S_ {\mu\nu}(k)
\end{equation}
where the index $j = 1,2,g$ denotes the twist-4 operators specified
in (\ref{eq:condt41}-\ref{eq:condt4g}).\footnote{The coefficients $A^j$ should 
not be confused with the parton distribution moments $A^i_n$.}

In \cite{choi93}
it is thoroughly discussed how the expectation values of these condensates 
with respect to single nucleon states
can be extracted from DIS experiments. To determine the contribution of these
condensates to the current-current correlator we follow the same strategy 
as described in subsection \ref{sec:t2s2}, i.e.,
\begin{equation}
  \label{eq:scattintro4}
\Pi^{d=6,\tau=4}_{\mu\nu}(q) 
= 4 \int\limits_{\vert \vec k \vert \le k_F } {d^3 k \over (2\pi)^3 \,2 E_k} 
T_{\mu\nu}^{s=2,\tau=4}(q,k)  
\end{equation}
with the twist-4 spin-2 contribution to the forward scattering amplitude between
a nucleon and the respective isospin current (\ref{eq:curud}),
\begin{equation}
  \label{eq:fsa24}
T_{\mu\nu}^{s=2,\tau=4}(q,k) := 
i \int\!\! d^4\!x \, e^{iqx} 
\langle N(\vec k) \vert 
T j_\mu(x) j_\nu(0) 
\vert N(\vec k) \rangle_{s=2,\tau=4}   \,,
\end{equation}
With the general ansatz (cf.~(\ref{eq:ans22}))
\begin{eqnarray}
T_{\mu\nu}^{s=2,\tau=4}(q,k) & = & 
\tilde B_1 \, [q^4 S_{\mu\nu}(k) - q^2 q_\mu q^\alpha S_{\nu\alpha}(k) 
- q^2 q_\nu q^\alpha S_{\mu\alpha}(k) 
+ g_{\mu\nu} q^2 q^\alpha q^\beta S_{\alpha\beta}(k)  ]
\nonumber \\
&& {} + \tilde B_2 \, (q_\mu q_\nu - g_{\mu\nu} q^2) \, 
q^\alpha q^\beta S_{\alpha\beta}(k) 
  \label{eq:ans24}
\end{eqnarray}
we get for the kinematical situation of DIS
\begin{equation}
  \label{eq:scattampdis24}
T_{\mu\nu}^{(\psi) \, s=2,\tau=2\, {\rm DIS}}(q,k) 
= - q^2 (q\cdot k)^2 (\tilde B_1 \, d_{\mu\nu} + \tilde B_2 \, e_{\mu\nu} )
\end{equation}
which has to be compared with eq.~(2) from \cite{choi93}. We end up with
\begin{equation}
  \label{eq:b1tilde}
\tilde B_1 = {4 \over q^8} \left( A^1 + {5 \over 8} A^2 + {1\over 16} A^g \right)
\end{equation}
and
\begin{equation}
  \label{eq:b2tilde}
\tilde B_2 = {4 \over q^8} \left( {1 \over 4} A^2 - {3\over 8} A^g \right)   \,.
\end{equation}
Based on a flavor decomposition new parameters $K^j_\psi$ are introduced in 
\cite{choi93} in terms of which the coefficients $A^j$ can be expressed. We refer
to \cite{choi93} for details and only give the final result for our isospin averaged
coefficients $A^j$:
\begin{equation}
  \label{eq:avk1}
A^1 = {1\over 4} [ K_u^1 + K_d^1 - (1 \pm 1) K_{ud}^1 ] \,,
\end{equation}
\begin{equation}
  \label{eq:avk2}
A^2 = {1\over 4} (K_u^2 + K_d^2)   \,,
\end{equation}
\begin{equation}
  \label{eq:avkg}
A^g = {1\over 4} (K_u^g + K_d^g)   \,,
\end{equation}
where the $+$ sign in (\ref{eq:avk1}) corresponds to the $\rho$ and the $-$ sign to 
the $\omega$ meson. The parameters $K^j_\psi$ are given by the expectation values
of the twist-4 condensates (\ref{eq:condt41}-\ref{eq:condt4g}) with respect to
a single proton state $p$ 
(cf.~(\ref{eq:condt41}-\ref{eq:condt4g},\ref{eq:trltwist4})):
\begin{equation}
  \label{eq:defku1}
\langle p(k) \vert {\cal ST} \, g^2 \, \bar u \gamma_\mu \gamma_5 \lambda^a u \,
( \bar u \gamma_\nu \gamma_5 \lambda^a u + \bar d \gamma_\nu \gamma_5 \lambda^a d)
\vert p(k) \rangle
=: K_u^1 S_ {\mu\nu}(k)   \,,
\end{equation}
\begin{equation}
  \label{eq:defku2}
\langle p(k) \vert {\cal ST} \, g^2 \,
\bar u \gamma_\mu \lambda^a u \, 
( \bar u \gamma_\nu \lambda^a u + \bar d \gamma_\nu \lambda^a d)
\vert p(k) \rangle
=: K_u^2 S_ {\mu\nu}(k)   \,,
\end{equation}
\begin{equation}
  \label{eq:defkug}
\langle p(k) \vert {\cal ST} \, ig \,
\bar u \, \{ D_\mu, \tilde G_{\nu\alpha} \} \gamma^\alpha \gamma_5 u  \,
\vert p(k) \rangle
=: K_u^g S_ {\mu\nu}(k)  
\end{equation}
and respective definitions for $K_d^j$, $j=1,2,g$. Furthermore,
\begin{equation}
  \label{eq:defkud1}
\langle p(k) \vert {\cal ST} \, g^2 \, 
\bar u \gamma_\mu \gamma_5 \lambda^a u \,
\bar d \gamma_\nu \gamma_5 \lambda^a d \,
\vert p(k) \rangle
=: K_{ud}^1 S_ {\mu\nu}(k)   \,.
\end{equation}
Since the flavor structure of $K_d^{1,2,g}$ and $K_u^{1,2,g}$ are governed
by the $d$ quark and the $u$ quark, respectively, it seems reasonable to assume
that the ratio is always the same \cite{choi93}:
\begin{equation}
  \label{eq:ratio12g}
{ K_d^1 \over K_u^1} = { K_d^2 \over K_u^2} = { K_d^g \over K_u^g} =: \beta \,.
\end{equation}
Within that assumption, the ratio $\beta$ and the quantities $\tilde B_1$ and 
$\tilde B_2$ for $\rho$ and $\omega$ can be determined from the DIS data. As 
expected from the valence quark decomposition of the proton one finds
\begin{equation}
  \label{eq:beta}
\beta \approx 0.5   
\end{equation}
within a small error. Furthermore we get 
\begin{equation}
  \label{eq:tb1rhom}
\tilde B_1 = {1 \over q^8} [18\,(c_1 - c_2) - (1 \pm 1) K_{ud}^1 ] 
\end{equation}
and
\begin{equation}
  \label{eq:tb2rhom}
\tilde B_2 = {1 \over q^8} 6 \, c_3   \,,
\end{equation}
where the constants $c_i$, $i$$=$$1,2,3$, as well as $K_{ud}^1$ can be found 
in table \ref{tab:para}. 

In total, the contribution of the twist-4 spin-2 operators is given by
(\ref{eq:scattintro4},\ref{eq:ans24},\ref{eq:tb1rhom},\ref{eq:tb2rhom}).
We note that a difference between the $\rho$ and
the $\omega$ meson within the OPE up to dimensionality $d$$=$$6$ only shows up for 
the twist-4 condensates and is 
expressed here in terms of the quantity $K_{ud}^1$. 

To summarize, we have presented in this section the operator product expansion
of the current-current correlator (\ref{eq:curcur}) including condensates up to
dimensionality $d$$=$$6$. The general form for the transverse and longitudinal part
of the current-current correlator which enter (\ref{eq:opehad}) is given by
\begin{equation}
  \label{eq:opegenform}
\Pi^{{\rm OPE}}_{T,L}(Q^2,\vec q\,^2) = Q^2 \, 
\left[
-{1 \over 8 \pi^2} \left( 1+ {\alpha_s \over \pi} \right) 
  {\rm ln}\left(Q^2\over \mu^2 \right) 
+ \sum\limits_{n}
{c_{T,L}^n(\vec q\,^2) \over Q^{2n}}   
\right]   \,,
\end{equation}
where the coefficients $c_{T,L}^n$ have to be deduced from the various contributions
discussed in this section. We note that contrary to the vacuum case 
\cite{shif79,leinw} and to the in-medium case with vanishing three-momentum 
$\vec q$ \cite{hats92,hats93,asako,hats95,jinlein,klingl97,leupold97} 
for the general case $\vec q \neq 0$ there are not only $1/Q^4$ and $1/Q^6$ terms
in (\ref{eq:opegenform}) but also higher order terms, even when we restricting
the OPE to $d$$\le$$6$ condensates. Appropriate powers of
$\vec q$ in the numerator serve to achieve the correct overall dimension. 
We will come back to that point in section \ref{sec:lindens}.

\section{QCD sum rule}  \label{sec:sumrule}

In the last section we have calculated the l.h.s.~of (\ref{eq:opehad}) within
the operator product expansion and some additional assumptions. We postpone the
discussion of these assumptions to section \ref{sec:crit} and present here some
general ideas about the calculation of the r.h.s.~of (\ref{eq:opehad}) and about
the use of this equation. 

If one has a model at hand which yields the current-current correlator
for arbitrary positive energy and arbitrary three-momentum, one could directly
use (\ref{eq:opehad}) to judge the reliability of this model. In practice, however,
the situation is such that one might have a model for the lowest hadronic resonance
in the respective isospin channel, i.e.~for $\rho$ and $\omega$, respectively, but 
one usually has no model which remains valid for arbitrary high energies. 
In the dispersion integral of (\ref{eq:opehad}) higher lying resonances are 
suppressed, but only by a factor $1/s^3$. Clearly, it is desirable to achieve a
larger suppression of the part of the hadronic spectral function on which one has 
less access. To this aim, a Borel transformation \cite{shif79,pastar} can be 
applied to (\ref{eq:opehad}). For an arbitrary function $f(Q^2)$ the Borel 
transformation is defined as
\begin{equation}
  \label{eq:fftildebo}
f(Q^2) \stackrel{\hat B}{\to} \tilde f(M^2)
\end{equation}
with 
\begin{equation}
  \label{eq:boreldef}
\hat B := \lim\limits_{{Q^2 \to \infty \,, \, N \to \infty 
                         \atop Q^2/N =: M^2 = {\rm fixed} }}
{1 \over \Gamma(N) } (-Q^2)^N \left( {d \over d Q^2} \right)^N   
\end{equation}
where $M$ is the so-called Borel mass. 

We will apply the Borel transformation to 
(cf.~(\ref{eq:opehad},\ref{eq:opegenform}))
\begin{eqnarray}
\lefteqn{-{1 \over 8 \pi^2} \left( 1+ {\alpha_s \over \pi} \right) 
  {\rm ln}\left(Q^2\over \mu^2 \right) 
+ \sum\limits_{n}
{c_{T,L}^n(\vec q\,^2) \over Q^{2n}}  } \nonumber \\
& = &
{\Pi^{{\rm HAD}}_{T,L}(0,\vec q\,^2) \over Q^2} - c_{T,L}(\vec q\,^2) 
+ {Q^2 \over \pi} \int\limits_{-\vecz{q}}^\infty \!\! ds \,
{{\rm Im}\Pi^{{\rm HAD}}_{T,L}(s,\vecz{q}) \over (s+Q^2-i\epsilon)(s+i\epsilon')^2} 
\,.  \label{eq:tbboreled}
\end{eqnarray}
Therefore we need to know the Borel transforms of $f(Q^2) = (Q^2+s)^{-\beta}$ and 
$f(Q^2) = {\rm ln} \, Q^2$. From the definition (\ref{eq:boreldef}) it is easy to
derive \cite{pastar}
\begin{equation}
  \label{eq:borat}
f(Q^2) = (Q^2+s)^{-\beta} \quad \Rightarrow \quad 
\tilde f(M^2) = {1\over \Gamma(\beta)} {1 \over M^{2\beta} } e^{-s/M^2}
\end{equation}
and
\begin{equation}
  \label{eq:boln}
f(Q^2) = {\rm ln} \, Q^2 \quad \Rightarrow \quad \tilde f(M^2) = -1   \,.
\end{equation}
Applying the Borel transformation to (\ref{eq:tbboreled}) we get 
\begin{eqnarray}
\lefteqn{{1 \over 8 \pi^2} \left( 1+ {\alpha_s \over \pi} \right) 
+ \sum\limits_{n} {1\over \Gamma(n)}
{c_{T,L}^n(\vec q\,^2) \over M^{2n}}  } \nonumber \\
& = &
{\Pi^{{\rm HAD}}_{T,L}(0,\vec q\,^2) \over M^2} 
- {1 \over \pi M^2} \int\limits_{-\vecz{q}}^\infty \!\! ds \,
{{\rm Im}\Pi^{{\rm HAD}}_{T,L}(s,\vecz{q}) \over s+i\epsilon} \, e^{-s/M^2}  \,.
  \label{eq:sumrules}
\end{eqnarray}
It is useful to write the r.h.s.~of the last equation in a form, where it is
more obvious that this expression is actually real valued. To this aim we split
$1/(s+i\epsilon)$ into a principal value and a $\delta$-function:
\begin{equation}
  \label{eq:princdel}
{1 \over s+i\epsilon} = {s \over s^2 + \epsilon^2 } - i \pi \delta(s)  \,.
\end{equation}
Using this decomposition we find the QCD sum rule
\begin{eqnarray}
\lefteqn{{1 \over 8 \pi^2} \left( 1+ {\alpha_s \over \pi} \right) 
+ \sum\limits_{n} {1\over \Gamma(n)}
{c_{T,L}^n(\vec q\,^2) \over M^{2n}}  } \nonumber \\
& = &
{{\rm Re}\Pi^{{\rm HAD}}_{T,L}(0,\vec q\,^2) \over M^2} 
- {1 \over \pi M^2} \int\limits_{-\vecz{q}}^\infty \!\! ds \,
{\rm Im}\Pi^{{\rm HAD}}_{T,L}(s,\vecz{q}) {s \over s^2+\epsilon^2} 
\, e^{-s/M^2}  \,.
  \label{eq:sumrule}
\end{eqnarray}
We observe that higher resonance states are now exponentially suppressed. 
Additionally we find a $1/s$ suppression. The latter is due to the fact that
we have applied the Borel transformation to $1/Q^2$ times equation (\ref{eq:opehad})
instead of directly applying it to (\ref{eq:opehad}). On the one hand, such
an additional suppression factor is desirable. On the other hand, we have to pay
a price for this, namely that the subtraction constant 
$\Pi^{{\rm HAD}}_{T,L}(0,\vec q\,^2)$ has not dropped out in contrast to the other
subtraction constant $c_{T,L}(\vec q\,^2)$. Had we applied the Borel transformation
to $1/Q^4$ times equation (\ref{eq:opehad}), the latter would also have survived. 
This is of course easy to understand from the point of view of subtracted dispersion
relations: the suppression of high energy contributions has to be compensated for
by a more detailed knowledge of the function at the subtraction point. 
We note that it is easy to get from (\ref{eq:sumrule}) also the direct Borel 
transformation of (\ref{eq:opehad}) without the $1/Q^2$ factor. We simply have
to multiply (\ref{eq:sumrule}) by $(-M^2)$ and differentiate with respect to $M^2$
afterwards. Using this recipe the subtraction constant 
$\Pi^{{\rm HAD}}_{T,L}(0,\vec q\,^2)$ obviously would drop out. Also the
$1/s$ suppression in the integral of (\ref{eq:sumrule}) would disappear. 

Having achieved a reasonable suppression of the energy region above the lowest
lying resonance the integral in (\ref{eq:sumrule}) is no longer sensitive to the
details of the hadronic spectral function in that region. For high energies
the quark structure of the current-current correlator is resolved. QCD perturbation
theory becomes applicable yielding
\begin{equation}
  \label{eq:qcdperthe}
{\rm Im}\Pi^{{\rm HAD}}_{T,L}(s,\vecz{q}) 
= - {s \over 8 \pi} \left( 1+ {\alpha_s \over \pi} \right)   \qquad 
\mbox{for large $s$.}
\end{equation}
These considerations suggest the ansatz
\begin{equation}
  \label{eq:pihadans}
{\rm Im}\Pi^{{\rm HAD}}_{T,L}(s,\vecz{q}) 
= \Theta(s_0 -s) \, {\rm Im}\Pi^{{\rm RES}}_{T,L}(s,\vecz{q}) 
+ \Theta(s -s_0) \, {-s \over 8 \pi} \left( 1+ {\alpha_s \over \pi} \right)   \,,
\end{equation}
where $s_0$ denotes the threshold between the low energy region described by a
spectral function for the lowest lying resonance, ${\rm Im}\Pi^{{\rm RES}}$, 
and the high energy region described by a continuum calculated from perturbative 
QCD. Of course, the
high energy behavior given in (\ref{eq:pihadans}) is only an 
approximation on the true spectral function for the current-current 
correlator. Also the rapid cross-over in (\ref{eq:pihadans}) from the resonance
to the continuum region is not realistic. However, exactly here the suppression
factors discussed above should become effective making a more detailed description
of the cross-over and the high energy region insignificant.

The price we have to pay for the simple decomposition (\ref{eq:pihadans}) is the
appearance of a new parameter $s_0$, the continuum threshold, which in general 
depends on the three-momentum $\vec q$ and on the nuclear density. We will elaborate
later on the determination of $s_0$. 

Inserting (\ref{eq:pihadans}) into (\ref{eq:sumrule}) yields
\begin{eqnarray}
\lefteqn{{1 \over 8 \pi^2} \left( 1+ {\alpha_s \over \pi} \right) \,
\left( 1 - e^{-s_0(\vecz{q})/M^2} \right)
+ \sum\limits_{n} {1\over \Gamma(n)}
{c_{T,L}^n(\vec q\,^2) \over M^{2n}}  } \nonumber \\
& = &
{{\rm Re}\Pi^{{\rm RES}}_{T,L}(0,\vec q\,^2) \over M^2} 
- {1 \over \pi M^2} \int\limits_{-\vecz{q}}^{s_0(\vecz{q})} \!\! ds \,
{\rm Im}\Pi^{{\rm RES}}_{T,L}(s,\vecz{q}) {s \over s^2 +\epsilon^2} 
\, e^{-s/M^2}  \,.
  \label{eq:sumrule2}
\end{eqnarray}
Obviously, the exponential suppression in (\ref{eq:sumrule2}) works only, if the
Borel mass $M$ is not too large. On the other hand, the OPE expression on the 
l.h.s.~of (\ref{eq:sumrule2}) gives a reliable prescription, if $M$ is not too 
small, since we have neglected higher order condensates which are accompanied by 
higher orders in $1/M^2$. At best, the sum rule (\ref{eq:sumrule2}) is valid inside
of a Borel window 
\begin{equation}
  \label{eq:borelwindow}
M^2_{\rm min} \leq M^2 \leq M^2_{\rm max}  \,,
\end{equation}
where $M^2_{\rm min}$ has to be determined such that the neglected condensates do
not spoil the validity of the l.h.s.~of (\ref{eq:sumrule2}), while  $M^2_{\rm max}$ 
has to be determined such that the suppression of the details in the high energy 
structure of the current-current correlator becomes effective. Of
course, it is not clear {\it a priori}, if such a Borel window exists at all. 
It might happen that $M^2_{\rm min}$ is larger than $M^2_{\rm max}$. 
In this worst case, the sum rule (\ref{eq:sumrule2}) would be useless.

The strategy to determine the Borel window is as follows \cite{leinw}: \\
--- For $M^2_{\rm min}$ we require that for this Borel mass the absolute value of
the contribution of the 
$d$$=$$6$ condensates is a certain percentage $p$ of the total absolute value of 
the l.h.s.~of (\ref{eq:sumrule}). 
Since the $d$$=$$6$ condensates have the highest order in mass which is
taken into account, one might expect that the relative contribution of the 
neglected condensates is much less than $p$.
Following \cite{leinw} we take $p = 10\%$, i.e.,
\begin{equation}
  \label{eq:detmmin}
\left\vert \sum\limits_{n} {1\over \Gamma(n)}
{c_{T,L}^{n,\, d=6}(\vec q\,^2) \over (M^2_{\rm min}){\vphantom{M^2}}^n} \right\vert
= 0.1 \, \left\vert
{1 \over 8 \pi^2} \left( 1+ {\alpha_s \over \pi} \right) 
+ \sum\limits_{n} {1\over \Gamma(n)}
{c_{T,L}^n(\vec q\,^2) \over (M^2_{\rm min}){\vphantom{M^2}}^n}
\right\vert   \,.
\end{equation}
--- For $M^2_{\rm max}$ we require that for this Borel mass the absolute value of
the continuum 
contribution to the integral in (\ref{eq:sumrule}) is a certain percentage $p'$
of the total absolute value of the integral. Again we follow \cite{leinw} and take 
$p' = 50\%$, i.e.,
\begin{equation}
  \label{eq:detmmax1}
\left\vert 
\int\limits_{-\vecz{q}}^\infty \!\! ds \, \Theta(s-s_0) \,
{\rm Im}\Pi^{{\rm HAD}}_{T,L}(s,\vecz{q}) {s \over s^2+\epsilon^2} 
\, e^{-s/M^2_{\rm max}} 
\right\vert = 0.5 \, \left\vert
\int\limits_{-\vecz{q}}^\infty \!\! ds \,
{\rm Im}\Pi^{{\rm HAD}}_{T,L}(s,\vecz{q}) {s \over s^2+\epsilon^2} 
\, e^{-s/M^2_{\rm max}} 
\right\vert    \,.
\end{equation}
By insertion of the decomposition (\ref{eq:pihadans}) we get 
\begin{equation}
  \label{eq:detmmax2}
{1 \over 8 \pi} \left( 1 + { \alpha_s \over \pi} \right) 
\, e^{-s_0(\vecz{q})/M^2_{\rm max}} 
= - \int\limits_{-\vecz{q}}^{s_0(\vecz{q})} \!\! ds \,
{\rm Im}\Pi^{{\rm RES}}_{T,L}(s,\vecz{q}) {s \over s^2+\epsilon^2} 
\, e^{-s/M^2_{\rm max}} 
\end{equation}
We note in passing that the sign of ${\rm Im}\Pi^{{\rm HAD}}_{T,L}$ and therefore
also the sign of ${\rm Im}\Pi^{{\rm RES}}_{T,L}$ is always negative 
(cf.~(\ref{eq:qcdperthe},\ref{eq:impiapl})). 

Obviously, the lower limit of the Borel window depends only on the condensates
calculated in section \ref{sec:ope}. In contrast to that, the upper limit 
depends on the choice for the continuum threshold $s_0$ and on the
hadronic model which yields $\Pi^{{\rm RES}}_{T,L}$. In general, both limits
may depend on the three-momentum $\vec q$ and on the nuclear density. 

\begin{figure}
\centerline{
\epsfxsize=\textwidth \epsfbox{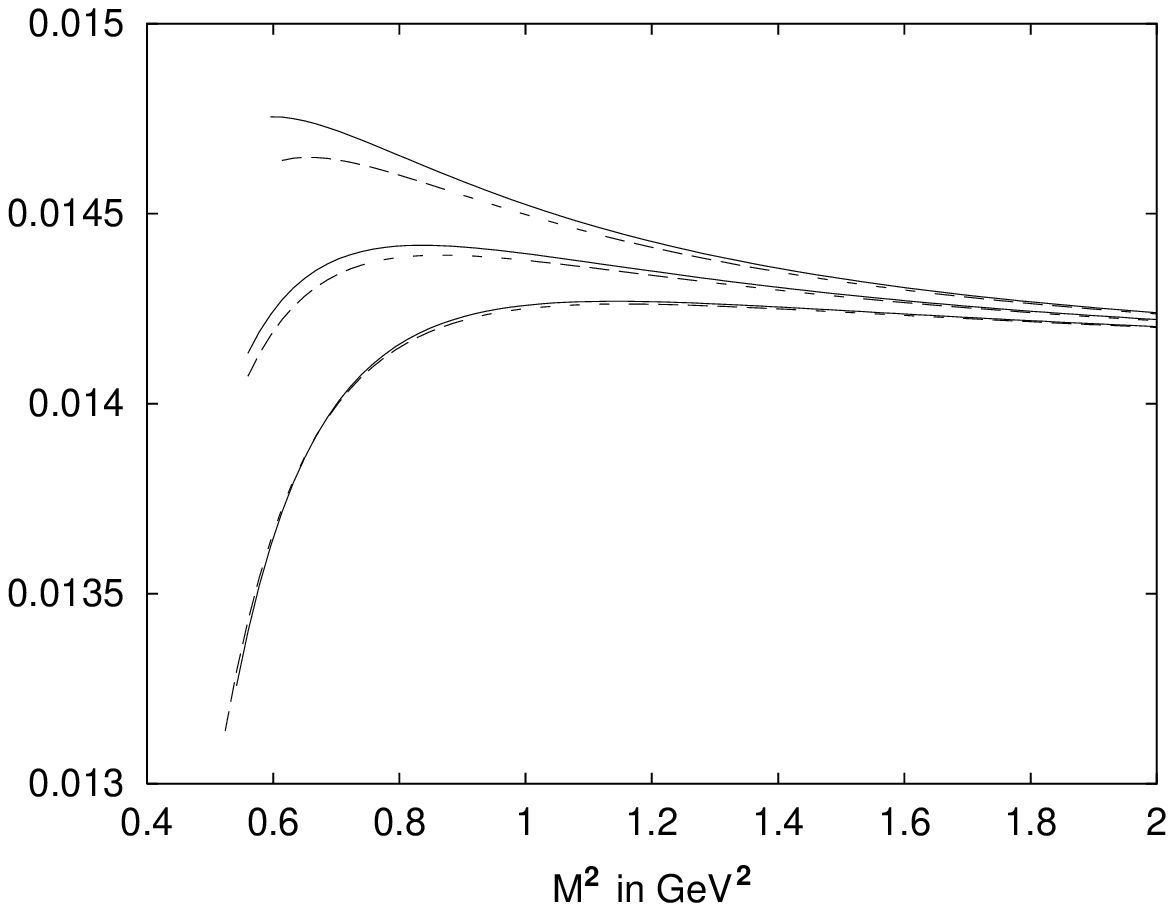}
}
\caption{Transverse part of l.h.s.~of (\protect\ref{eq:sumrule}) as a function
of the Borel mass squared, $M^2$, for 
three-momenta $\vert\vec q\vert = 0,\, 0.5,\, 1\,$GeV (top to bottom)
and for $\rho$ (full lines) and $\omega$
mesons (dashed lines).}  \label{fig:lhstrexact}
\end{figure}

\begin{figure}
\centerline{
\epsfxsize=\textwidth \epsfbox{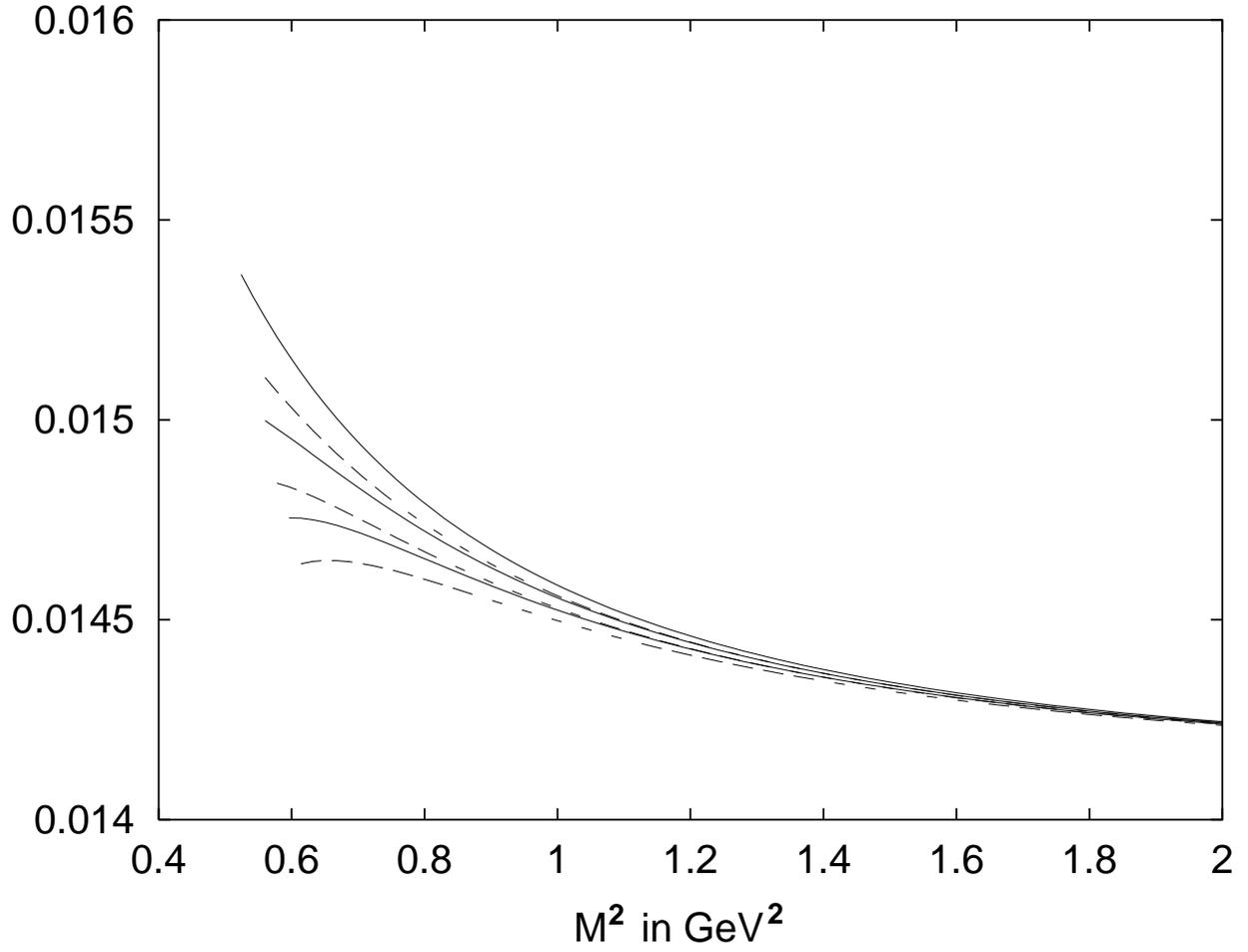}
}
\caption{Longitudinal part of l.h.s.~of (\protect\ref{eq:sumrule}) as a function
of the Borel mass squared, $M^2$, for 
three-momenta $\vert\vec q\vert = 0,\, 0.5,\, 1\,$GeV (bottom to top)
and for $\rho$ (full lines) and $\omega$
mesons (dashed lines).}  \label{fig:lhsloexact}
\end{figure}

Figure \ref{fig:lhstrexact} shows the transverse component of the l.h.s.~of 
(\ref{eq:sumrule}) as a function of the Borel mass squared, $M^2$, for various 
values of three-momentum squared, $\vecz{q}$, and for $\rho$ and $\omega$ meson. 
For the nuclear density we have chosen the nuclear saturation density of 
$0.17\,$fm$^{-3}$. 
Figure \ref{fig:lhsloexact} shows the same for the longitudinal component. 
On the left hand side both figures start with $M^2 = M^2_{\rm min}$ as deduced from
(\ref{eq:detmmin}). Obviously, the difference between $\rho$ and $\omega$ meson
is only very small and vanishes with rising $M^2$. The latter observation can
be easily understood recalling that the only difference in the OPE's for 
$\rho$ and $\omega$ comes from the twist-4 spin-2 condensates which are suppressed
at least by a factor $1/M^6$. Hence, the suppression becomes more effective with
rising $M^2$. Note that the small difference between $\rho$ and $\omega$ does not
necessarily mean that there is not much difference in their spectral functions. 
It only means that the integrated quantity given in (\ref{eq:sumrule2}) 
(to be rigorous: the r.h.s.~of that equation) is nearly
the same for both mesons. Nonetheless, the fact that the sum rule 
(\ref{eq:sumrule}) is nearly insensitive to the choice for the meson provides a 
strong constraint on a hadronic model which aims at a description of $\rho$ and
$\omega$ on the same footing, like e.g.~\cite{klingl97}.

We also observe that the dependence of the l.h.s.~of the
sum rule (\ref{eq:sumrule}) on the three-momentum $\vec q$ is rather weak. This
also constrains the hadronic model. Again, we stress that this does not mean that
the dependence of the spectral function on the three-momentum is weak. 

Suppose now that one has a hadronic model at hand which yields at least the 
imaginary part ${\rm Im}\Pi^{{\rm RES}}_{T,L}(s,\vecz{q})$ for the respective 
isospin channel at finite nuclear density. Examples can be found in 
\cite{chanfray,herrmann,asakawa,rapp1,friman,rapp2,klingl97,peters}. 
Then, for given nuclear density and three-momentum $\vec q$ one can utilize
the sum rule (\ref{eq:sumrule2}) and the results shown in figures 
\ref{fig:lhstrexact} and \ref{fig:lhsloexact} 
as a consistency check for the hadronic model
in the following way (see also \cite{leupold97}): 

\begin{itemize}
\item Choose a continuum threshold $s_0$ and a subtraction constant 
${\rm Re}\Pi^{{\rm RES}}_{T,L}(0,\vec q\,^2)$. 
\item Calculate the limits of the Borel window according to (\ref{eq:detmmin})
and (\ref{eq:detmmax2}). 
\item Calculate the relative deviation $r$ of the r.h.s.~from the l.h.s.~of the
sum rule (\ref{eq:sumrule2}), averaged over the Borel window, i.e.~schematically
\begin{equation}
\label{eq:diffdef}
  r = \int\limits^{M^2_{\rm max}}_{M^2_{\rm min}} \!\! d(M^2) \, 
\left\vert 1- {\rm r.h.s.}/ {\rm l.h.s.} \right\vert 
/ \Delta M^2   
\end{equation}
with
\begin{equation}
  \label{eq:delm}
  \Delta M^2 = M^2_{\rm max} - M^2_{\rm min}  \,.
\end{equation}
The l.h.s.~function can be taken from figures 
\ref{fig:lhstrexact} and \ref{fig:lhsloexact}.
\item Tune the ``fit parameters'' $s_0$ and 
${\rm Re}\Pi^{{\rm RES}}_{T,L}(0,\vec q\,^2)$ such that the deviation $r$ becomes
minimal. 
\end{itemize}
If this optimal $r$ is reasonably small and the Borel window not too small,
one might conclude that the considered hadronic model is in agreement with the QCD
sum rule for the chosen nuclear density and three-momentum $\vec q$. 

We close this section with some remarks on the respective size of the ``fit 
parameters'' $s_0$ and ${\rm Re}\Pi^{{\rm RES}}_{T,L}(0,\vec q\,^2)$. 
Clearly, the more fit parameters we have the less restrictive is the
sum rule for the hadronic model which should be checked. At least, it is therefore 
important to get an idea about the size and the possible influence of the fit 
parameters. 

In vacuum the continuum threshold $s_0$ turns out to be about
$1 - 1.6\,$GeV$^2$ \cite{leupold97,klingl97,leinw}. At least it 
has to be below the exited states of $\rho$ and $\omega$. Model calculations suggest
that the threshold decreases with increasing density 
\cite{hats92,jinlein,klingl97,leupold97}. 

Concerning the subtraction constant ${\rm Re}\Pi^{{\rm RES}}_{T,L}(0,\vec q\,^2)$
it is important to note that within the Fermi gas approximation it can be 
rigorously calculated for vanishing three-momentum \cite{hats95,klingl97}. Here
it turned out that it is so small that it would not change the results 
drastically, if it
is simply neglected. Unfortunately, the expectation that it could be neglected
also for finite three-momentum is presumably not justified. If we use the 
full electromagnetic current in (\ref{eq:curud}) instead of a part of it with
well-defined isospin, then within the Fermi gas approximation the transverse part
of the subtraction 
constant would simply be the real part of the
forward scattering amplitude $T(\vec q,\vec k)$ of 
a (real) photon with momentum $\vec q$ on a nucleon with momentum $\vec k$ 
averaged over the Fermi sphere, i.e.,
\begin{equation}
  \label{eq:emsub}
{\rm Re}\Pi^{{\rm em}}_T(0,\vec q\,^2) = 
4 \int\limits_{\vert \vec k \vert \le k_F } {d^3 k \over (2\pi)^3 \,2 E_k} \,
{\rm Re}T(\vec q,\vec k)  
\approx {\rho_N \over 2 m_N}  {\rm Re}T(\vec q,0)   \,,
\end{equation}
where we have used the linear density approximation (cf.~next section) to obtain 
the last expression.
It is reasonable to assume that ${\rm Re}\Pi^{{\rm RES}}_T(0,\vec q\,^2)$ is of
the same order of magnitude as ${\rm Re}\Pi^{{\rm em}}_T(0,\vec q\,^2)$. 
Thus, to get an idea about the size of the 
corresponding quantity in the sum rule (\ref{eq:sumrule2}) we have plotted in
figure \ref{fig:feuster} the quantity
\begin{equation}
  \label{eq:plotem}
{1 \over M^2} {\rho_N \over 2 m_N}  {\rm Re}T(\vec q,0) 
\end{equation}
as a function of the photon energy 
$E_\gamma = \vert \vec q \vert$ for nuclear saturation density and a typical
value for the Borel mass, $M = 1\,$GeV. We have used the model for photoproduction
presented in \cite{feuster98} to obtain the real part of the isospin averaged
photon-nucleon forward scattering amplitude. 
Comparing the absolute sizes in figure
\ref{fig:lhstrexact} and \ref{fig:feuster} we find indeed that the subtraction
constant might be negligible for $\vec q = 0$, but not for arbitrary 
three-momentum. Especially, if we are interested in the dependence on the
three-momentum we have to take the subtraction constant into account, since the
variation of the curves in \ref{fig:lhstrexact} with three-momentum is
of the same order of magnitude as the quantity plotted in figure \ref{fig:feuster}.
Concerning the longitudinal part of the subtraction constant we cannot compare
with photon-nucleon scattering, since there are no real longitudinal photons. 
Therefore, we refrain from presenting any estimates for this case. 

Of course, the most fortunate situation would be, if the considered hadronic
model already provides a value for ${\rm Re}\Pi^{{\rm RES}}_{T,L}(0,\vec q\,^2)$. 
Then, only the continuum threshold $s_0$ would remain as a fit parameter.

\section{Linear density approximation}   \label{sec:lindens}

Obviously, the contributions of the OPE presented in section \ref{sec:ope} are
quite unillustrative, simply due to its complexity. To get more insight in the
various contributions we restrict ourselves in this section to the parts which are 
at most linear in the nuclear density, i.e.~cubic in the Fermi momentum. Recalling
that we have evaluated all in-medium condensates using the Fermi gas approximation
(\ref{eq:fermi}) we have neglected nucleon-nucleon correlations anyway which
are quadratic in the density. Thus, the results presented in section \ref{sec:ope}
are at best correct up to Fermi momentum to the power of five, 
i.e.~up to $o(\rho_N^{5/3})$. Therefore, we do not lose too much information, if
we restrict ourselves here to the linear density approximation. To put it in 
physical terms: What we neglect further on is the Fermi motion of the nucleons. 
Anyway, for
concrete calculations we can use the full results presented above. We note in 
passing that actually all the required integrals over the Fermi sphere can be
calculated analytically. The results, however, are lengthy and unillustrative. 
Thus, for pedagogical reasons it is useful to discuss the linear density case,
\begin{equation}
  \label{eq:lindensappr}
4 \int\limits_{\vert \vec k \vert \le k_F } {d^3 k \over (2\pi)^3 \,2 E_k} 
\langle N(\vec k) \vert {\cal O} \vert N(\vec k) \rangle  
\to {\rho_N \over 2 m_N} \langle N(0) \vert {\cal O} \vert N(0) \rangle   \,.
\end{equation}
In this case, the coefficients $c_{T,L}^n(\vecz{q})$ introduced in 
(\ref{eq:opegenform}) are given by
\begin{equation}
  \label{eq:csplitd4d6}
c_{T,L}^n(\vecz{q}) = 
c_{T,L}^{n, \, d = 4}(\vecz{q}) + c_{T,L}^{n, \, d=6}(\vecz{q})
\end{equation}
with the contributions from the $d$$=$$4$ condensates to the transverse part,
\begin{eqnarray}
\lefteqn{c_T^{n=2, \, d = 4} = 
{1 \over 24} \left\langle {\alpha_s \over \pi} G^2 \right\rangle_0  
+ m_q \langle \bar q q \rangle_0}  \nonumber \\ && 
+ \rho_N \, \left\{ - {m_N^{(0)} \over 27} + {\sigma_N \over 2}  
+ {m_N \over 4} \left[   
A^{u+d}_2 (C^q_{2,2} -{3\over 2} C^q_{L,2}) 
+ {2\over 3} A^G_2 (C^G_{2,2} -{3\over 2}  C^G_{L,2} )
\right]
\right\}
\,,
\label{eq:ldc24t}
\end{eqnarray}
\begin{eqnarray}
\label{eq:ldc34t}
c_T^{n=3, \, d = 4}(\vecz{q}) = 
- \rho_N \vecz{q} {m_N \over 2} \left[ 
A_2^{u+d} (C_{2,2}^q - C_{L,2}^q) + {2 \over 3} A_2^G (C_{2,2}^G - C_{L,2}^G)
\right]   \,,
\end{eqnarray}
and to the longitudinal part,
\begin{eqnarray}
c_L^{n=2, \, d = 4} = c_T^{n=2, \, d = 4}   \,,
\label{eq:ldc24l}
\end{eqnarray}
\begin{eqnarray}
\label{eq:ldc34l}
c_L^{n=3, \, d = 4}(\vecz{q}) = 
\rho_N \vecz{q} {m_N \over 2} \left[ 
A^{u+d}_2 C^q_{L,2} + {2 \over 3} A_2^G C^G_{L,2} \right]   \,,
\end{eqnarray}
and the corresponding contributions from the $d$$=$$6$ condensates,
\begin{eqnarray}
\lefteqn{c_T^{n=3, \, d = 6} = 
-{112 \over 81} \kappa \alpha_s \pi \langle \bar q q \rangle_0^2   } \nonumber \\
&& + \rho_N \, \left\{ 
-{112 \over 81} {\sigma_N \over m_q} \kappa \alpha_s \pi \langle \bar q q \rangle_0
- { 5 m_N^3 \over 12} \left[ 
A_4^{u+d} (C_{2,4}^q - {3\over 2} C^q_{L,4} ) 
+ {2 \over 3} A_4^G (C_{2,4}^G - {3\over 2} C^G_{L,4} ) 
\right]
\right. \nonumber \\  && \left. \phantom{mmmm}
+ {9 m_N \over 2} \left[ 
c_1-c_2  - {1\over 2} c_3 - {1 \over 18} (1 \pm 1) K^1_{ud} 
\right]  \right\}
\,,
\label{eq:ldc36t}
\end{eqnarray}
\begin{eqnarray}
c_T^{n=4, \, d = 6}(\vecz{q}) & = &
\rho_N \vecz{q} \left\{ 
{9 m_N^3 \over 4} \left[
A_4^{u+d} (C_{2,4}^q - {10 \over 9} C_{L,4}^q) 
+ {2 \over 3} A_4^G (C_{2,4}^G - {10 \over 9} C_{L,4}^G) 
\right]
\right.  \nonumber \\ && \left. \phantom{mmm}
- 9 m_N \left[
c_1 - c_2 - {1 \over 3} c_3 - {1 \over 18} (1\pm 1) K_{ud}^1
\right]
\right\}
\,,
\label{eq:ldc46t}
\end{eqnarray}
\begin{eqnarray}
\label{eq:ldc56t}
c_T^{n=5, \, d = 6}(\vecz{q}) =
-2 \rho_N \vec q \, ^4 m_N^3 \left[
A_4^{u+d} (C_{2,4}^q - C_{L,4}^q) 
+{2 \over 3} A_4^G (C_{2,4}^G - C_{L,4}^G)  
\right]    \,,
\end{eqnarray}
and 
\begin{eqnarray}
\label{eq:ldc36l}
c_L^{n=3, \, d = 6} = c_T^{n=3, \, d = 6}   \,,
\end{eqnarray}
\begin{eqnarray}
\label{eq:ldc46l}
c_L^{n=4, \, d = 6}(\vecz{q}) = 
\rho_N \vecz{q} \left\{
{m_N^3 \over 2} \left[
A_4^{u+d} (C^q_{2,4} - 5 C^q_{L,4})
+ {2 \over 3} A_4^G (C^G_{2,4} - 5 C^G_{L,4})
\right]
+ 3 m_N c_3 
\right\}      \,,
\end{eqnarray}
\begin{eqnarray}
\label{eq:ldc56l}
c_L^{n=5, \, d = 6}(\vecz{q}) = 
2 \rho_N \vec q \, ^4 m_N^3 \left[
A_4^{u+d} C_{L,4}^q +{2 \over 3} A_4^G C_{L,4}^G  
\right]       \,.
\end{eqnarray}
All coefficients which are not given explicitly above vanish. As always, the 
$\pm$ sign which accompanies the $K_{ud}^1$ term corresponds to $\rho$ and $\omega$.

We immediately observe that the scalar condensates contribute only to the
three-momentum independent coefficients $c_{T,L}^{n=2,\,d=4}$ and
$c_{T,L}^{n=3,\,d=6}$. These coefficients are identical for transverse and
longitudinal part, since at $\vec q = 0$ we cannot distinguish between 
transverse and longitudinal directions \cite{hats93}. 

Obviously, there are contributions from numerous condensates to each of the 
coefficients presented above. 
On inspection of the parameters given in tables \ref{tab:para} and \ref{tab:coeff}
we can work out which condensates dominate which coefficient. Neglecting less
important condensates we find
\begin{equation}
  \label{eq:rough24}
c_T^{n=2, \, d = 4} = c_L^{n=2, \, d = 4} \approx 
{1 \over 24} \left\langle {\alpha_s \over \pi} G^2 \right\rangle_0  
+ \rho_N \,{m_N \over 4} \, A^{u+d}_2   \,,
\end{equation}
\begin{equation}
  \label{eq:rought34}
c_T^{n=3, \, d = 4}(\vecz{q}) \approx  
- \rho_N \vecz{q} {m_N \over 2} \, A_2^{u+d}  \,,
\end{equation}
\begin{equation}
  \label{eq:roughl34}
\vert c_L^{n=3, \, d = 4}(\vecz{q})  \vert \ll 
\vert c_T^{n=3, \, d = 4}(\vecz{q}) \vert   \,,
\end{equation}
\begin{equation}
  \label{eq:rought36}
c_T^{n=3, \, d = 6} = c_L^{n=3, \, d = 6} \approx 
-{112 \over 81} \kappa \alpha_s \pi \langle \bar q q \rangle_0^2 
- \rho_N \,
{112 \over 81} {\sigma_N \over m_q} \kappa \alpha_s \pi \langle \bar q q \rangle_0
\,,
\end{equation}
\begin{equation}
  \label{eq:rought46}
c_T^{n=4, \, d = 6}(\vecz{q}) \approx
\rho_N \vecz{q} \left\{ 
{9 m_N^3 \over 4} A_4^{u+d} 
- 9 m_N \left[
c_1 - c_2 - {1 \over 3} c_3 - {1 \over 18} (1\pm 1) K_{ud}^1
\right]
\right\}   \,,
\end{equation}
\begin{equation}
  \label{eq:rought56}
c_T^{n=5, \, d = 6}(\vecz{q}) \approx
-2 \rho_N \vec q \, ^4 m_N^3 A_4^{u+d}   \,,
\end{equation}
\begin{equation}
  \label{eq:roughl46}
  c_L^{n=4, \, d = 6}(\vecz{q}) \approx
\rho_N \vecz{q} \left(
{m_N^3 \over 2} \, A_4^{u+d} + 3 m_N c_3 
\right)   \,,
\end{equation}
\begin{equation}
  \label{eq:roughl56}
\vert c_L^{n=5, \, d = 6}(\vecz{q}) \vert \ll
\vert c_T^{n=5, \, d = 6}(\vecz{q}) \vert   \,.
\end{equation}
Especially we have neglected all $\alpha_s$ corrections to the twist-2 condensates,
i.e.~we have approximated $C^q_{2,n}$ by 1 and neglected all other $C^i_{r,n}$
(cf.~\cite{flor81}). For vanishing three-momentum $\vec q$ the density dependent
terms are dominated by the twist-2 spin-2 quark condensate and the scalar 
four-quark condensate. Concerning the $\vecz{q}$-terms it is remarkable that
the twist-4 condensates are equally important as the twist-2 spin-4 quark 
condensates. Of course, both are suppressed for large Borel masses as compared to 
the $(d$$=$$4)$ terms. Thus, for the transverse part the twist-2 spin-2 quark
condensate governs the $\vecz{q}$-terms. For the longitudinal direction the 
situation is more involved, since the $(d$$=$$4)$ coefficient given in 
(\ref{eq:ldc34l}) is quite small. Therefore, we have competing contributions from
(\ref{eq:ldc34l}) and (\ref{eq:roughl46}). 

We stress that in principle it is not necessary to perform the approximations 
which have led from (\ref{eq:ldc24t}-\ref{eq:ldc56l}) to 
(\ref{eq:rough24}-\ref{eq:roughl56}). Of course, one can use the exact expressions
for the coefficients. The purpose here was only to figure out the condensates which
have the most influence on the coefficients. 

Since we have non-vanishing coefficients up to $n$$=$$5$ we find contributions
up to $o(1/Q^{10})$. One may suspect that it is inconsistent to keep terms
of order $o(1/Q^8)$ and higher,
since we have neglected ($d$$=$$8$) condensates which would contribute at
$o(1/Q^8)$. However, this is misleading, since the dependence on the three-momentum
$\vec q$ is different in both cases. Schematically the neglected higher order 
condensates would contribute as 
\begin{equation}
  \label{eq:connegqdep}
{(d=8) \, \mbox{condensate} \over Q^8} \times \left( \# + \# {\vecz{q} \over Q^2}
+ \# {\vec q \,^4 \over Q^4} + \dots  \right)
\end{equation}
where $\#$ denotes arbitrary dimensionless numbers which do not depend on $q$.
Thus, e.g.~the $\vecz{q}$-terms of the neglected condensates are actually 
$o(1/Q^{10})$, while the corresponding terms of the condensates taken into account 
are $o(1/Q^8)$. 

To work out the dependence of the current-current correlator on the
three-momentum $\vec q$ more explicitly we study (\ref{eq:sumrule}) in the vicinity
of $\vec q = 0$. For vanishing three-momentum we find
\begin{eqnarray}
\lefteqn{{1 \over 8 \pi^2} \left( 1+ {\alpha_s \over \pi} \right) 
+ {c^{n=2, \, d=4} \over M^4} + {c^{n=3, \, d=6} \over 2 M^6}   } \nonumber \\
& = &
{{\rm Re}\Pi^{{\rm HAD}}(0,0) \over M^2} 
- {1 \over \pi M^2} \int\limits_{0}^\infty \!\! ds \,
{\rm Im}\Pi^{{\rm HAD}}(s,0) {s \over s^2+\epsilon^2} 
\, e^{-s/M^2}  \,.
  \label{eq:sumruleq0}
\end{eqnarray}
We have skipped the label $T,L$, since there are no distinct directions at 
vanishing three-momentum. 
Next we differentiate (\ref{eq:sumrule}) with respect to $\vecz{q}$ and put
$\vec q = 0$ afterwards. This yields
\begin{eqnarray}
\lefteqn{
{d_{T,L}^{n=3,\, d=4} \over 2 M^6} + {d_{T,L}^{n=4,\, d=6} \over 6 M^8}
} \nonumber \\
& = &  {d \over d (\vecz{q}) } \left. \left(
{{\rm Re}\Pi^{{\rm HAD}}_{T,L}(0,\vec q\,^2) \over M^2} 
- {1 \over \pi M^2} \int\limits_{-\vecz{q}}^\infty \!\! ds \,
{\rm Im}\Pi^{{\rm HAD}}_{T,L}(s,\vecz{q}) {s \over s^2+\epsilon^2} 
\, e^{-s/M^2}  
\right) \right\vert_{\vec q =0} 
  \label{eq:sumruledq0}
\end{eqnarray}
with 
\begin{equation}
  \label{eq:defdc}
d_{T,L}^{n=3,\, d=4} = {c_{T,L}^{n=3,\, d=4}(\vecz{q}) \over \vecz{q}} 
\quad , \quad 
d_{T,L}^{n=4,\, d=6} = {c_{T,L}^{n=4,\, d=6}(\vecz{q}) \over \vecz{q}} \,.
\end{equation}
In the same way we find by differentiating twice:
\begin{equation}
{d_{T,L}^{n=5,\, d=6} \over 24 M^{10}}
= {d^2 \over d (\vecz{q})^2 } \left. \left(
{{\rm Re}\Pi^{{\rm HAD}}_{T,L}(0,\vec q\,^2) \over M^2} 
- {1 \over \pi M^2} \int\limits_{-\vecz{q}}^\infty \!\! ds \,
{\rm Im}\Pi^{{\rm HAD}}_{T,L}(s,\vecz{q}) {s \over s^2+\epsilon^2} 
\, e^{-s/M^2}  
\right) \right\vert_{\vec q =0} 
  \label{eq:sumruledtq0}
\end{equation}
with 
\begin{equation}
  \label{eq:defdc56}
d_{T,L}^{n=5,\, d=6} = {c_{T,L}^{n=5,\, d=6}(\vecz{q}) \over \vec{q}\,^4} \,.
\end{equation}
Let us discuss now the range of validity for the new sum rules 
(\ref{eq:sumruledq0}) and (\ref{eq:sumruledtq0}). As pointed out in section 
\ref{sec:ope} it is crucial to find a non-vanishing Borel window where the 
validity of the sum rule is guaranteed inside of this window. To find the lower
limit of this window we have compared the contribution of the highest order 
condensates to the l.h.s.~of (\ref{eq:sumrule}) with the total result 
(cf.~(\ref{eq:detmmin})). In (\ref{eq:sumruleq0}) we have four different orders in
$1/M^2$, namely zeroth to third order.\footnote{Note that the first order term
vanishes. Strictly speaking it is proportional to the light current quark mass 
squared which is negligibly small.} Thus, it is no problem to compare the
third order contribution to the total result. In (\ref{eq:sumruledq0}) we are
left with third and fourth order in $1/M^2$, only. Thus, the number of orders
we have access on is already diminished. This leads to a lower limit of the Borel
window of $M^2_{\rm min} \approx 3\,$GeV$^2$ for the transverse component of 
sum rule (\ref{eq:sumruledq0}) which is already much higher 
than the one for sum rule (\ref{eq:sumruleq0}): 
$M^2_{\rm min} \approx 0.6\, $GeV$^2$. For the longitudinal component of 
sum rule (\ref{eq:sumruledq0}) we even find $M^2_{\rm min} \approx 10\,$GeV$^2$. 
As already discussed above the $\vecz{q}$-part of the $(d$$=$$4)$ contribution
to the longitudinal part (\ref{eq:ldc34l}) is quite small. Therefore, only for 
very large values of the Borel mass the $(d$$=$$4)$ contribution can overwhelm 
the $(d$$=$$6)$ contribution. Both for the longitudinal and the transverse
part we find that the respective lower limit of the Borel window for sum rule 
(\ref{eq:sumruledq0}) is much higher than the one for (\ref{eq:sumruleq0}). 
If the
upper limit of the Borel window does not rise in the same way, the 
sum rule (\ref{eq:sumruledq0}) would not be as useful as (\ref{eq:sumruleq0}). To
determine the upper limit of the Borel window we would need, of course, a hadronic
model. For the most simple case, the approximation of the spectral function by a
$\delta$-function, it was found in \cite{lee97} that the upper limit of the Borel
window also rises. Thus, also the sum rule (\ref{eq:sumruledq0}) might be useful
as a consistency check for hadronic models. 
We note, however, that the definition of the Borel window in \cite{lee97}
differs from ours. 
For (\ref{eq:sumruledtq0}) the situation is even worse. There we have only access 
on one order in $1/M^2$. Thus, we cannot determine a lower limit for the
Borel window. The sum rule (\ref{eq:sumruledtq0}) is therefore useless. 

We stress again that the approximations performed in this section are not 
mandatory. The purpose of these approximations was to obtain more qualitative
insight in the importance of the various contributions and in the dependence on
the three-momentum $\vec q$. 
To check the consistency of a hadronic model with the QCD sum rule
(\ref{eq:sumrule2}) the OPE coefficients should be deduced from the formulae
presented in section \ref{sec:ope}. Only if the hadronic model is also restricted
to the linear density case, a direct comparison with the simplified expressions
would be appropriate.

\section{Discussion of the assumptions}  \label{sec:crit}

In this section we will discuss the various assumptions that have led to
the condensate contributions calculated in section \ref{sec:ope}. 

The basic 
assumption we have made for the evaluation of in-medium condensates is the 
Fermi gas approximation (\ref{eq:fermi}). Clearly, this approximation is only
valid for not too high densities. Of course, it would be of interest to quantify 
this statement. Unfortunately this is hard to do within the OPE approach, since
it is not clear how to calculate expectation values with respect to multi-nucleon
states for arbitrary nuclear densities. An idea about the importance of 
multi-nucleon states might be obtained from a comparison of the moments of parton 
distributions (\ref{eq:defmomq},\ref{eq:defmomg}) as deduced from DIS experiments
with nucleons on the one hand side and with nuclei on the other. This might be a
direction of future studies. Concerning the parton distributions in nuclei we
refer to \cite{eskola98} and references therein.

Within the framework of a hadronic model it is tedious but possible to
approximately take into account interactions of the vector meson with more than 
one nucleon \cite{chanfray,herrmann,asakawa,rapp1,friman,rapp2,peters}. In
\cite{peters} it was found that already at nuclear saturation density it is
important to account for such processes. On the other hand, presumably
in every hadronic model one can distinguish between single and multi-nucleon 
interactions. Therefore, it should be possible to compare the OPE calculation
with the hadronic model treated in a ``single nucleon mode''. This comparison was
e.g.~performed in \cite{klingl97} for vanishing three-momentum $\vec q$. 
If in the framework of a hadronic model it turns out that interactions with
more than one nucleon are important for the nuclear density under consideration,
then the QCD sum rule cannot serve to check the consistency of the whole 
hadronic model but only of its restriction to scattering processes of the vector
meson with one nucleon from the Fermi sphere. 

After these general considerations about the validity of the calculation of 
in-medium condensates we turn now to the discussion of the accuracy in the 
determination of the different types of condensates. 

Concerning the scalar condensates of subsection \ref{sec:scal} we have neglected
there terms which are quadratic in the light current quark masses $m_q$ as well as 
possible differences in the condensates of up and down quarks. In view of the
fact that the light current quark masses are about $6\,$MeV, while the Borel 
window for the QCD sum rule (\ref{eq:sumrule}) starts at about 
$M^2_{\rm min} \approx 0.6\,$GeV$^2$
(cf.~figures \ref{fig:lhstrexact}, \ref{fig:lhsloexact}), the neglect of 
$m_q^2$ terms is very well justified. Also, a possible difference between 
up and down quark condensate is presumably small and anyway hard to disentangle
in view of the uncertainties in the determination of the (average) light quark
condensate (see e.g.~discussions in \cite{leupold97,leinw}). In addition, the
contribution from the two-quark condensate to (\ref{eq:scal3}) is much smaller 
than the one from the gluon condensate. The largest
uncertainty lies in the evaluation of the four-quark condensate, i.e.~in the value
for $\kappa$. First, even the vacuum value is still under discussion. Second, it
might very well be that the value for $\kappa$ varies with nuclear density.
Concerning the first problem, it is useful to choose a hadronic model which 
describes the data for $e^+\,e^- \to $ hadrons reasonably well and utilize the
sum rule (\ref{eq:sumrule2}) for the vacuum case to determine $\kappa$. This was
performed in \cite{klingl97} and we therefore have taken the condensate values given
there. For the second problem we have no solution to offer so far. Without any 
better knowledge we use the vacuum value for $\kappa$ also at finite density. 
To get rid of this uncertainty one can differentiate the sum rule (\ref{eq:sumrule})
with respect to the three-momentum squared $\vecz{q}$. In this way, all scalar
condensates drop out, since they do not yield $\vecz{q}$-dependent contributions
to the OPE side of the sum rule. As already discussed in section \ref{sec:lindens}
this differentiated sum rule results in a lower limit for the Borel window
which is much higher than the one for the original sum rule. This might diminish
or even close the Borel window, i.e.~the thus obtained sum rule might be less
reliable or even useless. This clearly depends on the explicit hadronic model under
consideration. 

The twist-2 condensates discussed in subsections \ref{sec:t2s2}, \ref{sec:t2s4}
are the best known contributions, as soon as one accepts the Fermi gas approximation
discussed above. We even can use the results of DIS to get an idea about the
neglected higher order condensates (see below). 

The twist-4 spin-2 condensates can in principle also be deduced from DIS data. 
The uncertainties in their extraction are, however, quite large. We have adopted
the analysis of \cite{choi93}. There, condensates depending on the light current
quark masses, strange quark contributions and dependences on the renormalization
scale were neglected. These errors are presumably smaller than the uncertainties
in the extraction of these condensates from DIS data. In general, the contributions
of the twist-4 spin-2 condensates are small as compared to the twist-2 spin-2
contributions. For the $\vecz{q}$-contributions to the longitudinal direction,
however, the twist-2 spin-2 contribution is suppressed by $\alpha_s$. There, the
twist-4 spin-2 condensates cannot be disregarded \cite{lee97}. 

Of course, one is forced to somewhere cut off the OPE used here as an expansion in
the dimensionality of the condensates. We have neglected all condensates 
of dimensionality 8 or higher. 
To get an idea about the size of the neglected condensates we calculate now the
twist-2 spin-6 contribution to the current-current correlator, since we have
access on that quantity utilizing the DIS results. The calculation proceeds
along the lines described in subsections \ref{sec:t2s2}, \ref{sec:t2s4}. 
We use the Fermi gas approximation
\begin{equation}
  \label{eq:fermispin6}
\langle {\cal O}_{\mu\nu\kappa\lambda\xi\chi} \rangle \approx 
4 \int\limits_{\vert \vec k \vert \le k_F } {d^3 k \over (2\pi)^3 \,2 E_k} 
\langle N(\vec k) \vert 
{\cal O}_{\mu\nu\kappa\lambda\xi\chi} 
\vert N(\vec k) \rangle  
\end{equation}
and the decomposition
\begin{eqnarray}
\lefteqn{\langle N(\vec k) \vert {\cal O}_{\mu\nu\kappa\lambda\xi\chi} 
\vert N(\vec k) \rangle} \nonumber \\  
& \sim & 
k_\mu k_\nu k_\kappa k_\lambda k_\xi k_\chi
- {1\over 12}( k_\mu k_\nu k_\kappa k_\lambda g_{\xi\chi} m_N^2 
+ \mbox{14 permutations})
\nonumber \\ && {}
+ {1\over 120} (k_\mu k_\nu g_{\kappa\lambda} g_{\xi\chi} m_N^4 
+ \mbox{44 permutations}) 
- {1\over 960} (g_{\mu\nu} g_{\kappa\lambda} g_{\xi\chi} m_N^6 
+ \mbox{14 permutations})
\nonumber \\ 
&=:&  S_{\mu\nu\kappa\lambda\xi\chi}(k)   \,.
  \label{eq:trlspin6}
\end{eqnarray}
The relevant operators are 
\begin{equation}
  \label{eq:op26q}
{\cal ST} \,i( \bar u \gamma_\mu D_\nu D_\kappa D_\lambda D_\xi D_\chi u 
+ \bar d \gamma_\mu D_\nu D_\kappa D_\lambda D_\xi D_\chi d )
\end{equation}
and
\begin{equation}
  \label{eq:op26g}
{\cal ST} \, G^\alpha_{\phantom{\alpha}\mu} D_\nu D_\kappa D_\lambda D_\xi 
G_{\alpha\chi}   \,.
\end{equation}
From the DIS calculations we can deduce the following contribution to the 
current-current correlator (\ref{eq:curcur}) which was neglected in 
(\ref{eq:curcurdtau}):
\begin{equation}
  \label{eq:piscam26fin}
\Pi^{d=8,\tau=2}_{\mu\nu}(q) 
= \int\limits_{\vert \vec k \vert \le k_F } {d^3 k \over (2\pi)^3 \,2 E_k} 
T_{\mu\nu}^{(u+d) \, s=6,\tau=2}(q,k)  
\end{equation}
with the forward scattering amplitude
\begin{eqnarray}
T_{\mu\nu}^{(u+d) \, s=6,\tau=2}(q,k) & = & -{64 \over q^{14}} 
(q_\mu q_\nu q^\alpha q^\beta - g_\mu^{\phantom{\mu} \alpha} q^2 q_\nu q^\beta
- g_\nu^{\phantom{\nu} \alpha} q^2 q_\mu q^\beta 
+ g_\mu^{\phantom{\mu} \alpha} g_\nu^{\phantom{\nu} \beta} q^4 ) \, 
q^\gamma q^\delta q^\epsilon q^\zeta 
S_{\alpha \beta \gamma \delta \epsilon \zeta}(k) 
\nonumber \\
&& \phantom{-} \times 
\left( C_{2,6}^q A_4^{u+d} + {2\over 3} C_{2,6}^G A_4^G \right)
\nonumber \\
&&
-{64 \over q^{14}} (q_\mu q_\nu - g_{\mu\nu} q^2) \, 
q^\alpha q^\beta q^\gamma q^\delta q^\epsilon q^\zeta 
S_{\alpha \beta \gamma \delta \epsilon \zeta}(k) 
\nonumber \\
&& \phantom{-} \times 
\left( (C_{L,6}^q - C_{2,6}^q) A_4^{u+d} 
+ {2\over 3} (C_{L,6}^G - C_{2,6}^G) A_4^G \right)   \,.
  \label{eq:scam26fin}
\end{eqnarray}
With this at hand we can calculate 
the ratio between the twist-2 spin-6
contribution to the l.h.s.~of the sum rule (\ref{eq:sumrule}) and the total
value for this l.h.s.~as calculated in section \ref{sec:ope}. 
We have calculated that ratio in figures \ref{fig:errortr}, \ref{fig:errorlo} for
transverse and longitudinal directions and for $\rho$ and $\omega$ meson. 
Obviously, the obtained ratios are very small justifying at least the neglect
of twist-2 spin-6 condensates and suggesting that all higher dimensional condensates
are reasonably suppressed. 
Of course, for a thorough discussion of the error made by neglecting higher
dimensional condensates we also would need to know the other condensates beside
the twist-2 spin-6 condensates. Since we do not know e.g.~the scalar $(d$$=$$8)$
condensates etc.~the error analysis presented here is only a first guess. 

As long as we do not know the actual values of the neglected condensates (or at 
least an upper limit for them) we cannot present a rigorous proof that the 
OPE approach works, i.e.~that the truncated series yields a reliable value for
the current-current correlator in the region of interest. Indeed, in 
\cite{eletsky} it was doubted that the QCD sum rule approach provides any reliable
information about the medium modifications of vector mesons
(see also \cite{reply1,reply2}). Two arguments
were given there to support these doubts: The first, qualitative argument concerns
the connection between the mass shift of a vector meson and the 
forward scattering amplitude of this vector meson with a nucleon. It was argued
there that the forward scattering amplitude and hence also the mass shift is a
long distance property, while the OPE is only capable of short distance properties.
We think that this argument is misleading, since the OPE always concerns the
interplay of long and short distance properties, as already pointed out in section 
\ref{sec:ope}. Actually, in the same oversimplifying way one might argue just 
the other way round: The
in-medium $\rho$ mass is still large, i.e.~a short distance property, and therefore 
can be described by the OPE approach. This shows that one needs more
quantitative arguments to check the validity of the sum rule approach. 
A useful selfconsistency check within the Borel sum rule method is the 
Borel stability analysis described in section \ref{sec:sumrule}: A breakdown
of the sum rule might be observed in a small or even vanishing Borel window. 
Indeed, this stability analysis was the key point to resolve the 
question, whether mass shift and/or forward scattering amplitude 
can be extracted 
within the traditional sum rule approach utilizing the narrow width approximation.
We refer to \cite{koike95,hats95,jinlein,koike97} for details. In general, the
Borel window can only be determined after specifying a hadronic model.
Therefore, we do not discuss this point here any further. The preceding discussion
has clearly emphasized the necessity to perform a Borel stability analysis. 

A second, formal argument has been raised in \cite{eletsky} against the 
applicability of the OPE approach to vector mesons in nuclear medium: It was
claimed there that the OPE turns out to be an expansion in the nucleon mass
$m_N$ over the invariant mass $\sqrt{Q^2}$. After Borelization this would turn
into an expansion in $m_N$ over the Borel mass $M$. If the latter is assumed to 
be of the order of the $\rho$ meson mass, one would get an expansion parameter 
$m_N/m_\rho$ which is obviously not small. Therefore, it was argued in 
\cite{eletsky} that the truncation of the OPE at the $d$$=$$6$ condensates is
not appropriate. Indeed, concerning the twist-2 condensates the statement is
true that the used OPE is an expansion in $m_N/M$. This can most easily be discussed
within the linear density approximation of section \ref{sec:lindens}. E.g.~for
vanishing three-momentum the twist-2 spin-2 condensates contribute with a term
proportional to 
$\rho_N m_N/M^4$ (cf.~(\ref{eq:ldc24t})) and the twist-2 spin-4 condensates
with a term proportional to $\rho_N m_N^3/M^6$ (cf.~(\ref{eq:ldc36t})).  
In general, the twist-2 spin-$s$ condensates yield a contribution
$\sim \rho_N m_N^{s-1}/M^{s+2}$. Since the nucleon mass $m_N$ is large this
expansion might break down for the Borel masses of interest (typically of the
order of $1\,$GeV). However, one should not discuss the convergence of a series
without looking at its coefficients. 
The twist-2 spin-$s$ contribution is accompanied by 
the $s$-th moment of the parton distributions. Inspecting the last two lines
of table \ref{tab:coeff} we find that these moments become very small with 
increasing $s$. Indeed, we have already discussed above that the twist-2 spin-6
condensates do not spoil the truncation of the OPE --- in spite of the fact that
they are proportional to $\rho_N m_N^5 /M^8$. This shows that also the second
argument raised in \cite{eletsky} against the QCD sum rule approach is 
oversimplified.

\section{Summary}  \label{sec:summary}

This work was motivated by the finding that the QCD sum rule approach provides
no model independent prediction about a possible mass shift of vector mesons in 
nuclear medium \cite{klingl97,leupold97}. In \cite{leupold97} we have discussed
at length that the sum rule restricts the
$\rho$ meson only to a rather wide area in the (mass, width) plane without making 
any further statements about the specific properties of the $\rho$ meson. 
Only within additional model assumptions the behavior of the $\rho$ meson in 
nuclear matter can be further specified. E.g., if one {\it assumes} that the width
of the $\rho$ meson is not increased, then the sum rule predicts a 
$\rho$ mass which decreases with increasing nuclear density. However, it is 
also possible to {\it assume} instead that the $\rho$ mass is not shifted. In
this case the sum rule suggests an increasing width of the spectral function of the
$\rho$ meson. 

This, however, does not mean that
the sum rule approach is useless:
We have presented here a QCD sum rule for $\rho$ and $\omega$ mesons
propagating with arbitrary three-momentum through nuclear matter at vanishing 
temperature. This sum rule provides a non-trivial consistency check for hadronic 
models which describe that propagation. At least as long as different hadronic
models cannot be judged unambiguously by experiments such consistency
checks are important to confirm or rule out hadronic models. 

The main formula is given in (\ref{eq:sumrule2}). The OPE coefficients 
$c_{T,L}^n$ which
appear on the l.h.s.~of this formula are defined via $\Pi^{\rm OPE}_{T,L}$ in
(\ref{eq:opegenform}). In view of their complexity we have not given the 
explicit formulae for $c_{T,L}^n$. However, they can be easily deduced 
in the following way from the
equations presented in section \ref{sec:ope}: Transverse
and longitudinal part of $\Pi^{\rm OPE}_{T,L}$ are obtained from the respective last
expression of (\ref{eq:pit}) and (\ref{eq:pil}). The current-current correlator
with the full Lorentz structure, $\Pi^{\rm OPE}_{\mu\nu}$ is decomposed in 
(\ref{eq:curcurdtau}). 

The scalar contribution is given in (\ref{eq:scalope1})
where $R^{\rm scalar}$ can be read off from (\ref{eq:scal3}). The expectation
values showing up there are decomposed in vacuum and medium dependent 
expectation values
in (\ref{eq:condscal}) using the Fermi gas approximation. The vacuum expectation
values are listed in table \ref{tab:para}. Finally, the medium dependent parts of
the scalar condensates are connected in (\ref{eq:nucsig},\ref{eq:traceano}) with
parameters also listed in table \ref{tab:para}. 

The contribution from the twist-2 spin-2 condensates is given in 
(\ref{eq:piscam22fin},\ref{eq:scam22fin}) with the traceless tensor 
$S_{\alpha\beta}$ defined in (\ref{eq:trls}). In the same way the 
contribution from the twist-2 spin-4 condensates is given in 
(\ref{eq:piscam24fin},\ref{eq:scam24fin}) with the traceless tensor 
$S_{\alpha\beta\gamma\delta}$ defined in (\ref{eq:trlspin4}). 
The required values for the moments of the
parton distributions $A^i_n$ and the coefficients $C^i_{r,n}$ are collected in
table \ref{tab:coeff}. 

The contribution from the twist-4 spin-2 condensates is given in 
(\ref{eq:scattintro4},\ref{eq:ans24}) where the coefficients $\tilde B_{1,2}$
are connected in (\ref{eq:tb1rhom},\ref{eq:tb2rhom}) to quantities listed in
table \ref{tab:para}. 

In this way, the OPE coefficients can be easily calculated.
As discussed in the last section the QCD sum rule (\ref{eq:sumrule2}) 
can be used to check the consistency of a hadronic model, provided that in the 
latter the medium is also described by the Fermi gas approximation. Going one
step further by neglecting the Fermi motion of the nucleons both in the hadronic
model under consideration and in the calculation of the OPE coefficients 
one might also utilize the sum rule in this linear density approximation. For this
case the OPE coefficients $c_{T,L}^n$ are explicitly given in 
(\ref{eq:csplitd4d6}-\ref{eq:ldc56l}). 

By inspecting the QCD sum rule (\ref{eq:sumrule2}) we observe that the hadronic
model which should be checked has to yield the current-current correlator 
for invariant masses in the region $(-\vecz{q})$ to $s_0$. The lower limit refers
to vanishing energy. For non-vanishing three-momentum $\vec q$ this means that we
need information not only about the time like, but also about the space like region.
For small three-momenta the space like region is dominated by the coupling of the
respective vector meson to nucleon-hole states \cite{hats95}. For higher 
three-momenta also resonance-hole loops come into play in the space like region 
(see figure 3 in \cite{peters}). Thus, at finite nuclear density there are 
important structures in the
spectral function of the vector mesons also in the space like region. This is 
in contrast to the vacuum case where there is no structure below the two-pion
(three-pion) threshold for the $\rho$ ($\omega$) meson. 

Qualitatively, we have found that our sum rules are not very sensitive to the
difference between $\rho$ and $\omega$ meson as well as to a variation in 
the three-momentum of the vector meson with respect to the nuclear medium. This,
however, does not a priori mean that the current-current 
correlator for different isospin
channels and for different three-momenta should be more or less the same. In the sum
rule only an integral over this correlator enters which might be the same e.g.~for
$\rho$ and $\omega$ mesons, even if the respective correlators themselves are
different. Thus, on
this qualitative level the sum rule approach does not rule out hadronic models
which predict a different behavior of vector mesons with different three-momenta,
like e.g.~\cite{friman,rapp2,eletsky,peters,kondrat98}. A quantitative analysis
of these models is beyond the scope of this paper. 

We believe that the QCD sum rule presented here provides an interesting and
non-trivial consistency check for hadronic models which describe vector mesons
in nuclear matter. We have tried to present the derivation of the sum rule 
in great detail to make it possible also for non-experts in OPE to utilize the
sum rule for a consistency check of their hadronic models.

\begin{figure}
\centerline{
\epsfxsize=\textwidth \epsfbox{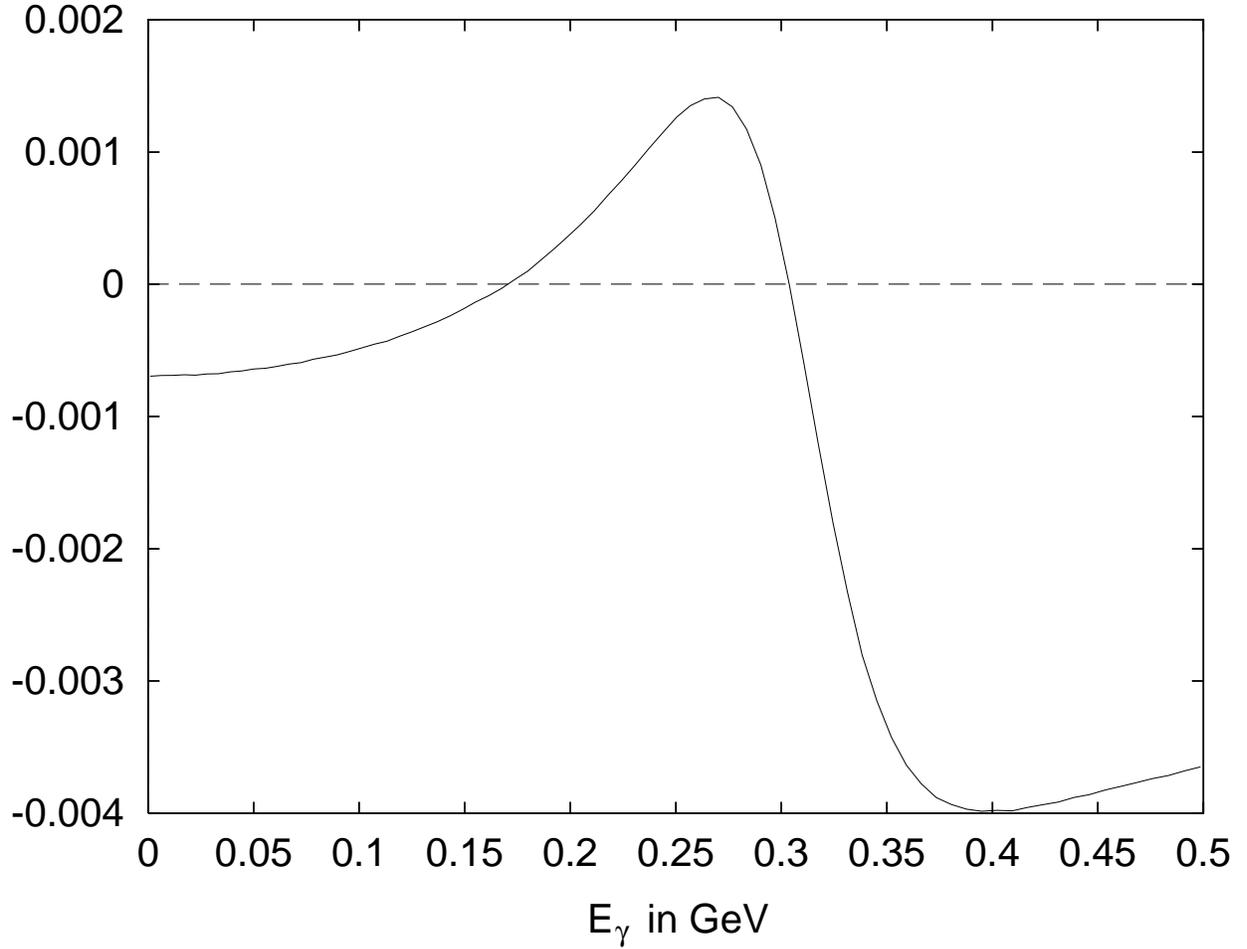}
}
\caption{Real part of isospin averaged
$\gamma N$ forward scattering amplitude as a function of the photon 
energy in the rest frame of the nucleon (rescaled with $\rho_N/(2m_N M^2)$, see
main text for details).} \label{fig:feuster}
\end{figure}

\begin{figure}
\centerline{
\epsfxsize=\textwidth \epsfbox{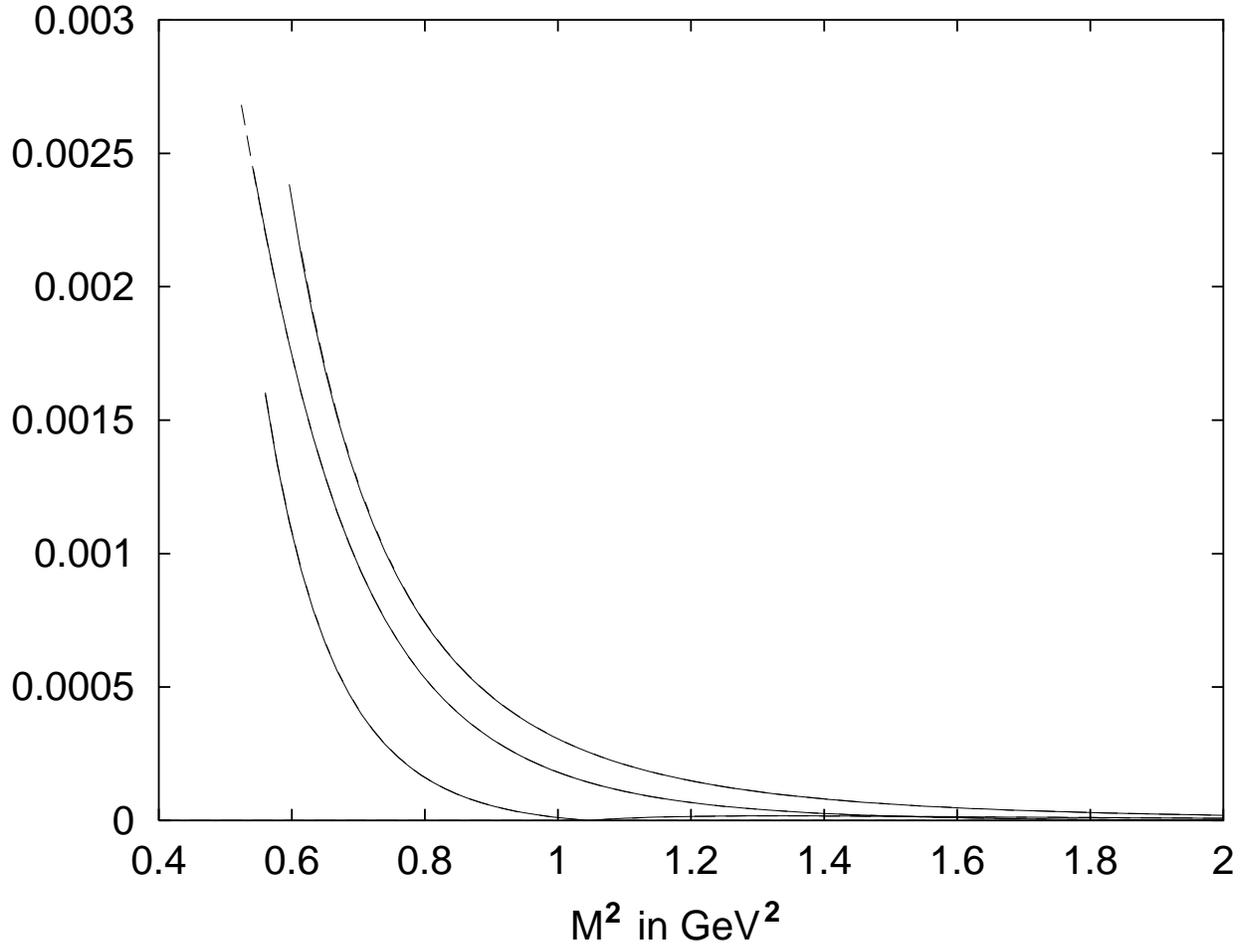}
}
\caption{Relative error made by the neglect of twist-2 spin-6 condensates in the
calculation of the 
transverse part of the l.h.s.~of (\protect\ref{eq:sumrule}) as a function
of the Borel mass squared, $M^2$, for 
three-momenta $\vert\vec q\vert = 0$ (upper line), 0.5 (lower line), $1\,$GeV 
(middle line)
and for $\rho$ (full lines) and $\omega$
mesons (dashed lines). The only noticeable difference between $\rho$ and $\omega$ 
meson appears 
in the slightly different lower limits of the Borel window, i.e.~in the
starting points of the curves on the left hand side.} \label{fig:errortr}
\end{figure}

\begin{figure}
\centerline{
\epsfxsize=\textwidth \epsfbox{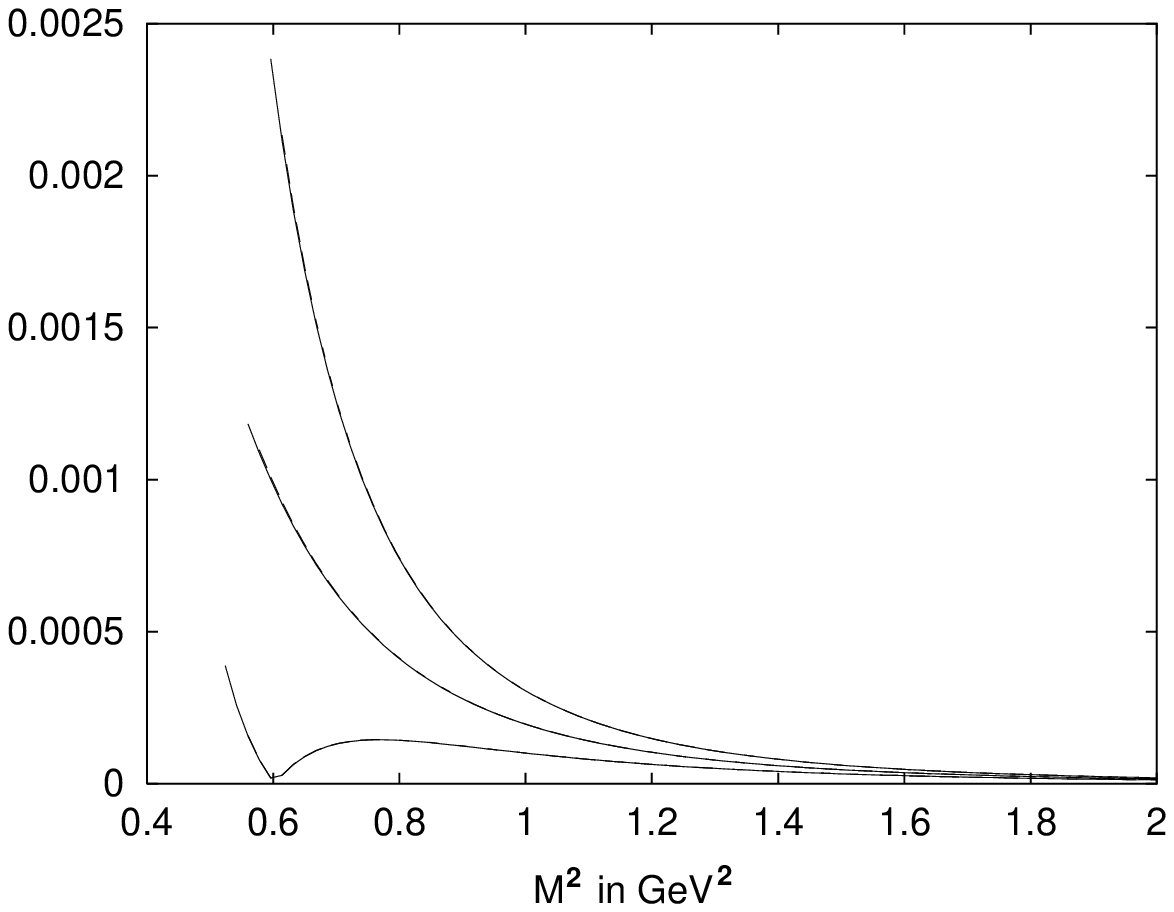}
}
\caption{Same as figure \protect\ref{fig:errortr} for longitudinal part. The 
curves refer to $\vert\vec q\vert = 0,\, 0.5,\, 1\,$GeV (top to bottom).}
\label{fig:errorlo}
\end{figure}

\acknowledgments
One of the authors (SL) acknowledge helpful discussions with S.H.~Lee, F.~Klingl, 
W.~Peters, M.~Post, F.~de Jong, and C.~Greiner. The authors also 
thank T.~Feuster for providing them with the data of figure \ref{fig:feuster}.

\begin{table}[ht]
\begin{tabular}{r|r||r|r||r|r}
$\alpha_s$ & 0.36  & $\sigma_N$ $[$GeV$]$ & 0.045 & $c_1$ $[$GeV$^2]$ & 0.005 \\
$m_q$ $[$GeV$]$ & 0.006 & $m_N^{(0)}$ $[$GeV$]$ & 0.75 
& $c_2$ $[$GeV$^2]$ & 0.011 \\
$\langle \bar q q \rangle_0$ $[$GeV$^3]$ & -0.0156 & & & $c_3$ $[$GeV$^2]$ & 0.035\\
$\langle (\alpha_s/\pi) G^2 \rangle_0$ $[$GeV$^4]$ & 0.012 & & 
& $K_{ud}^1$ $[$GeV$^2]$ & -0.088 \\
$\kappa$ & 2.36 & & & & 
\end{tabular}
\caption{Parameters used in the calculation of the OPE
contributions to the current-current correlator. See main text for details. }
\label{tab:para} 
\end{table}

\begin{table}[ht]
\begin{tabular}{l|c||l|c||l|c}
$C^q_{2,2}$ & 1.013 & $C^q_{2,4}$ & 1.171 & $C^q_{2,6}$ & 1.316 \\
$C^q_{L,2}$ & 0.050 & $C^q_{L,4}$ & 0.030 & $C^q_{L,6}$ & 0.022 \\
$C^G_{2,2}$ & -0.042 & $C^G_{2,4}$ & -0.063 & $C^G_{2,6}$ & -0.060 \\
$C^G_{L,2}$ & 0.057 & $C^G_{L,4}$ & 0.023 & $C^G_{L,6}$ & 0.012 \\
\hline \hline
$A^{u+d}_2$ & 1.12 & $A^{u+d}_4$ & 0.11 & $A^{u+d}_6$ & 0.03 \\
$A^G_2$ & 0.83 & $A^G_4$ & 0.04 & $A^G_6$ & 0.01 
\end{tabular}
\caption{Relevant coefficient functions $C^j_{i,n}$ taken from 
\protect\cite{flor81} and moments of parton distributions $A^j_n$ calculated from 
\protect\cite{glu92} for $\mu^2 = 1\,$GeV$^2$.} \label{tab:coeff}
\end{table}


\begin{references}
\bibitem{ceres1} G.~Agakichiev et al., CERES collaboration, 
Phys.~Rev.~Lett. {\bf 75} (1995) 1272. 
\bibitem{ceres2} A.~Drees for the CERES collaboration, in Proc.~of the International
Workshop XXIII on Gross Properties of Nuclei and Nuclear Excitations, 
Hirschegg 1995, eds.~H.~Feldmeier and W.~N\"orenberg, (GSI Darmstadt 195), p.143. 
\bibitem{ceres3} N.~Masera for the HELIOS-3 collaboration, 
Nucl.~Phys. {\bf A590} (1995) 93c. 
\bibitem{brownrho} G.E.~Brown and M.~Rho, Phys.~Rev.~Lett. {\bf 66} (1991) 2720. 
\bibitem{cassing} W.~Cassing, W.~Ehehalt, and C.M.~Ko, 
        Phys.~Lett. {\bf B363} (1995) 35.
\bibitem{ko} G.Q.~Li, C.M.~Ko, and G.E.~Brown, 
        Phys.~Rev.~Lett. {\bf 75} (1995) 4007;
        Nucl.~Phys. {\bf A606} (1996) 568.
\bibitem{bratkov} E.L.~Bratkovskaya and W.~Cassing, Nucl.~Phys. {\bf A619} 
(1997), 413. 
\bibitem{pisarski} R.~Pisarski, 
        Phys.~Rev. {\bf D52} (1995) R3773. 
\bibitem{chanfray} G.~Chanfray and P.~Schuck, 
Nucl.~Phys. {\bf A545} (1992) 271c; 
Nucl.~Phys. {\bf A555} (1993) 32.
\bibitem{herrmann} M.~Herrmann, B.~Friman and W.~N\"orenberg,
Nucl.~Phys. {\bf A545} (1992) 267c; Nucl.~Phys. {\bf A560} (1993) 411. 
\bibitem{asakawa} M.~Asakawa, C.M.~Ko, P.~L\'evai, and X.J.~Qiu, 
Phys.~Rev. {\bf C46} (1992) R1159. 
\bibitem{rapp1} R.~Rapp, G.~Chanfray, and J.~Wambach, 
Phys.~Rev.~Lett. {\bf 76} (1996) 368. 
\bibitem{friman} B.~Friman and H.J.~Pirner, Nucl.~Phys. {\bf A617} (1997) 496. 
\bibitem{rapp2} R.~Rapp, G.~Chanfray, and J.~Wambach, 
Nucl.~Phys. {\bf A617} (1997) 472.
\bibitem{klingl97} F.~Klingl, N.~Kaiser, and W.~Weise, Nucl.~Phys. {\bf A624} 
(1997) 527.
\bibitem{eletsky} V.L.~Eletsky and B.L.~Ioffe, Phys.~Rev.~Lett. {\bf 78} (1997) 
1010.
\bibitem{peters} W.~Peters, M.~Post, H.~Lenske, S.~Leupold, and U.~Mosel, 
Nucl.~Phys. {\bf A632} (1998) 109.
\bibitem{kondrat98} L.A.~Kondratyuk, A.~Sibirtsev, W.~Cassing, Ye.S.~Golubeva, 
and M.~Effenberger, nucl-th/9801055.
\bibitem{brat98} W.~Cassing, E.L.~Bratkovskaya, R.~Rapp, and J.~Wambach, 
Phys.~Rev. {\bf C57} (1998) 916.
\bibitem{shif79} M.A.~Shifman, A.I.~Vainshtein, and V.I.~Zakharov, 
 Nucl.~Phys. {\bf B147} (1979) 385, 448. 
\bibitem{reinders85} L.J.~Reinders, H.~Rubinstein, and S.~Yazaki, 
Phys.~Rep. {\bf 127} (1985) 1.
\bibitem{leinw} D.B.~Leinweber, Ann.~Phys. {\bf 254} (1997) 328. 
\bibitem{hats93} T.~Hatsuda, Y.~Koike, and S.H.~Lee, 
 Nucl.~Phys. {\bf B394} (1993) 221. 
\bibitem{adami} C.~Adami and I.~Zahed, Phys.~Rev. {\bf D43} (1991) 921.
\bibitem{hats92} T.~Hatsuda and S.H.~Lee, Phys.~Rev. {\bf C46} (1992) R34. 
\bibitem{hats95} T.~Hatsuda, S.H.~Lee, and H.~Shiomi, Phys.~Rev. {\bf C52} (1995)
3364. 
\bibitem{cohen95} T.D.~Cohen, R.J.~Furnstahl, D.K.~Griegel, X.~Jin,  
Prog.~Part.~Nucl.~Phys. {\bf 35} (1995) 221. 
\bibitem{jinlein} X.~Jin and D.B.~Leinweber, Phys.~Rev. {\bf C52} (1995) 3344. 
\bibitem{lee97} S.H.~Lee, Phys.~Rev. {\bf C57} (1998) 927.
\bibitem{leupold97} S.~Leupold, W.~Peters, and U.~Mosel, Nucl.~Phys. {\bf A628}
(1998) 311. 
\bibitem{gale91} C.~Gale and J.~Kapusta, Nucl.~Phys. {\bf B357} (1991) 65.
\bibitem{KB} L.P.~Kadanoff and G.~Baym, `Quantum Statistical Mechanics',
Benjamin, New York (1962).
\bibitem{wilson69} K.G.~Wilson, Phys.~Rev.~{179} (1969) 1499. 
\bibitem{bard78} W.A.~Bardeen, A.J.~Buras, D.W.~Duke, and T.~Muta, Phys.~Rev. 
{\bf D18} (1978) 3998.
\bibitem{flor81} E.G.~Floratos and C.~Kounnas, Nucl.~Phys. {\bf B192} (1981) 417.
\bibitem{glu92} M.~Gl\"uck, E.~Reya, and A.~Vogt, Z.~Phys. {\bf C53} (1992) 127.
\bibitem{pastar} P.~Pascual and R.~Tarrach, QCD: Renormalization for the
Practitioner, Lecture Notes in Physics, Vol. 194 (Springer, Berlin, 1984). 
\bibitem{shur81} E.V.~Shuryak and A.I.~Vainshtein, Phys.~Lett. {\bf B105} (1981) 65;
Nucl.~Phys. {\bf B199} (1982) 451.
\bibitem{jaffe81} R.L.~Jaffe and M.~Soldate, Phys.~Lett. {\bf B105} (1981) 467;
Phys.~Rev. {\bf D26} (1982) 49.
\bibitem{choi93} S.~Choi, T.~Hatsuda, Y.~Koike, and S.H.~Lee, Phys.~Lett. {\bf B312}
(1993) 351.
\bibitem{lee94} S.H.~Lee, Phys.~Rev. {\bf D49} (1994) 2242.
\bibitem{asako} M.~Asakawa and C.M.~Ko, Phys.~Rev. {\bf C48} (1993) R526. 
\bibitem{feuster98} T.~Feuster and U.~Mosel, nucl-th/9803057.
\bibitem{eskola98} K.J.~Eskola, V.J.~Kolhinen, and P.V.~Ruuskanen, hep-ph/9802350.
\bibitem{reply1} T.~Hatsuda and S.H.~Lee, nucl-th/973022.
\bibitem{reply2} V.L.~Eletsky and B.L.~Ioffe, hep-ph/9704236.
\bibitem{koike95} Y.~Koike, Phys.~Rev. {\bf C51} (1995) 1488.
\bibitem{koike97} Y.~Koike, A.~Hayashigaki, Prog.~Theor.~Phys. {\bf 98} (1997) 631.

\end{references}
\end{document}